\documentclass[twocolumn]{aastex631}

\newcommand{\msol}{$M_\odot$}
\newcommand{\vxhi}{\bar{x}_{\rm HI}}
\newcommand{\logtq}{\log{t_{\rm Q}}}
\newcommand{\zbar}{\langle z \rangle }
\newcommand{\chimp}{{\rm c Mpc}/h}

\newcommand{\lya}{Ly$\alpha$\ }
\newcommand{\lyb}{Ly$\beta$\ }

\newcommand{\cii}{[C{\small II}]}
\newcommand{\civ}{C{\small IV}}
\newcommand{\mgii}{Mg{\small II}}
\newcommand{\oiii}{[O{\small III}]}

\newcommand{\change}[1]{#1}

\usepackage{amsmath,amssymb,amsfonts}
\usepackage[caption=false]{subfig}
\usepackage[capitalise]{cleveref}

\begin{document}

\title{Chronicling the reionization history at $6\lesssim z \lesssim 7$ with emergent quasar damping wings}

\author[0000-0001-8986-5235]{Dominika {\v D}urov{\v c}{\'i}kov{\'a}}
\affiliation{MIT Kavli Institute for Astrophysics and Space Research, 77 Massachusetts Avenue, Cambridge, 02139, Massachusetts, USA}
\affiliation{Department of Physics, Massachusetts Institute of Technology, 77 Massachusetts Avenue Cambridge, MA 02139}

\author[0000-0003-2895-6218]{Anna-Christina Eilers}
\affiliation{MIT Kavli Institute for Astrophysics and Space Research, 77 Massachusetts Avenue, Cambridge, 02139, Massachusetts, USA}

\author[0000-0002-3211-9642]{Huanqing Chen}
\affiliation{Canadian Institute for Theoretical Astrophysics, University of Toronto,60 St George St, Toronto, ON M5R 2M8, Canada}

\author[0000-0001-5818-6838]{Sindhu Satyavolu}
\affiliation{Tata Institute of Fundamental Research, Homi Bhabha Road, Mumbai 400005, India}

\author[0000-0001-5829-4716]{Girish Kulkarni}
\affiliation{Tata Institute of Fundamental Research, Homi Bhabha Road, Mumbai 400005, India}

\author[0000-0003-3769-9559]{Robert A.\ Simcoe}
\affiliation{MIT Kavli Institute for Astrophysics and Space Research, 77 Massachusetts Avenue, Cambridge, 02139, Massachusetts, USA}

\author[0000-0001-5211-1958]{Laura C.\ Keating}
\affiliation{Institute for Astronomy, University of Edinburgh, Blackford Hill, Edinburgh, EH9 3HJ, UK}

\author[0000-0001-8443-2393]{Martin G.\ Haehnelt}
\affiliation{Institute of Astronomy, University of Cambridge, Madingley Road, Cambridge CB3 0HA, UK}
\affiliation{Kavli Institute of Cosmology, University of Cambridge, Madingley Road, Cambridge CB3 0HA, UK}

\author[0000-0002-2931-7824]{Eduardo Ba{\~n}ados}
\affiliation{Max Planck Institut f\"ur Astronomie, K\"onigstuhl 17, D-69117, Heidelberg, Germany}

\correspondingauthor{Dominika {\v D}urov{\v c}{\'i}kov{\'a}}
\email{dominika@mit.edu}



\begin{abstract}

The spectra of high-redshift ($z\gtrsim 6$) quasars contain valuable information on the progression of the Epoch of Reionization (EoR). At redshifts $z<6$, the observed Lyman-series forest shows that the intergalactic medium (IGM) is nearly ionized, while at $z>7$ the observed quasar damping wings indicate high neutral gas fractions. However, there remains a gap in neutral gas fraction constraints at $6\lesssim z \lesssim 7$ where the Lyman series forest becomes saturated but damping wings have yet to fully emerge.
In this work, we use a sample of 18 quasar spectra at redshifts $6.0<z<7.1$ to close this gap. We apply neural networks to reconstruct the quasars' continuum emission around the partially absorbed Lyman $\alpha$ line to normalize their spectra, and stack these continuum-normalized spectra in three redshift bins. To increase the robustness of our results, we compare the stacks to a grid of models from two hydrodynamical simulations, ATON and CROC, and we measure the volume-averaged neutral gas fraction, $\vxhi$, while jointly fitting for the mean quasar lifetime, $t_{\rm Q}$, for each stacked spectrum. We chronicle the evolution of neutral gas fraction using the ATON (CROC) models as follows: $\vxhi = 0.21_{-0.07}^{+0.17}$ ($\vxhi = 0.10_{<10^{-4}}^{+0.73}$) at $\zbar =6.10$, $\vxhi = 0.21_{-0.07}^{+0.33}$ ($\vxhi =0.57_{-0.47}^{+0.26}$) at $\zbar =6.46$, and $\vxhi = 0.37_{-0.17}^{+0.17}$ ($\vxhi =0.57_{-0.21}^{+0.26}$) at $\zbar =6.87$. At the same time we constrain the average quasar lifetime to be $t_{\rm Q} \lesssim 7\ {\rm Myr}$ across all redshift bins, in good agreement with previous studies.

\end{abstract}

\keywords{}


\section{Introduction} \label{sec:intro}

The Epoch of Reionization (EoR) represents the last major phase transition of our Universe, where the hydrogen in the intergalactic medium (IGM) transitions from a completely neutral state to the ionized state we observe today. The past decade has brought crucial insights into the ionizing sources (e.g. \citet{robertson_galaxy_2022}), as well as the morphology and timing (e.g. \citet{bosman_hydrogen_2022}) of the reionization process, all of which are important to understand open questions about galaxy formation, cosmology, and the evolution of the earliest supermassive black holes. Despite this progress, many details of the EoR remain unknown.

The timing of the EoR in particular has been studied extensively. For instance, the reionization optical depth inferred from the cosmic microwave background (CMB) favours a late and fast reionization and yields a midpoint of $z_\mathrm{re}\gtrsim 7$ \citep{planck_collaboration_planck_2020}. At redshifts $z\lesssim 6$, the observed Lyman $\alpha$ and $\beta$ forests in quasar spectra show that the \change{IGM is still not fully ionized} \citep{fan_constraining_2006,mcgreer_model-independent_2015,bosman_new_2018,eilers_opacity_2018,eilers_anomaly_2019,zhu_long_2022,bosman_hydrogen_2022}. 
However, the \lya absorption saturates already at relatively low volume averaged neutral gas fractions of $\vxhi \sim 10^{-4}$, leading to complete flux absorption and the formation of a Gunn-Peterson \change{trough} \citep{gunn_density_1965}. On the other hand, high neutral gas fractions at redshifts around $z\gtrsim 7$ can be probed via off-resonant (damped) absorption features in quasar spectra, i.e. damping wings \citep{miralda-escude_reionization_1998} imprinted on the red side of the \lya emission line. \change{As damping wings require a very high neutral hydrogen column density to be observable, current measurements} are limited to a few, high-redshift quasar sightlines \citep{banados_800-million-solar-mass_2018,greig_ly_2017,greig_constraints_2019,greig_igm_2022,davies_quantitative_2018,durovcikova_reionization_2020,yang_poniuaena_2020,wang_significantly_2020,mesinger_constraints_2007,schroeder_evidence_2013}\change{, indicating high neutral gas fractions of $\vxhi \gtrsim 30\%$. Securely modeling damping wings for lower $\vxhi$ along individual sightlines is precluded by the potential weakness of the damping wing signal combined with} uncertainties in the predictions of quasar continua. 
This leaves us with a range of neutral gas fraction spanning a few orders of magnitude that is challenging to probe with either the Lyman-series forest or the damping wing signature. Thus, there remains a gap in $\vxhi$ constraints at intermediate neutral gas fractions (corresponding roughly to $6 \lesssim z \lesssim 7$) where most quasars do not show strong damping wings \change{and are therefore not usable for single-sightline modeling}. 

Current upper bounds on the neutral hydrogen fraction at $6 \lesssim z \lesssim 7$ come from the \lya and \lyb forest dark pixel counts \citep{jin_nearly_2023,zhu_probing_2023} and are consistent with constraints from the CMB. In addition, galaxy-based constraints at $z\gtrsim 6.5$ are beginning to emerge in the literature. Lyman Break galaxies (LBGs) \citep{mason_universe_2018,mason_inferences_2019}, Lyman $\alpha$ emitters (LAEs) \change{\citep{ouchi_statistics_2010,sobacchi_clustering_2015,ning_magellan_2022}} as well as stacked galaxy damping wings \citep{umeda_jwst_2023} all point towards a significant neutral gas fraction, $\vxhi \gtrsim \change{30}\% $, at $z\gtrsim 6.5$.

This work aims to fill this gap in the history of the EoR. We do so by tackling two main challenges. First, \change{the insensitivity of the damping wing signature to many orders of magnitude in $\vxhi$ combined with the patchiness of reionization \citep[e.g.][]{iliev_simulating_2006,pentericci_new_2014,becker_reionisation_2015,bosman_new_2018,kulkarni_large_2019}} renders single-sightline quasar studies challenging near the end of the EoR, where \change{damping wings are either exceptionally strong or not visible at all.} Second, the quasar's radiation field affects the surrounding IGM. Quasar activity is accompanied by the release of radiation that ionizes the surrounding IGM over time, creating a region known as the proximity zone, thereby modifying the observed spectrum in the vicinity of the \lya line \change{\citep[e.g.][]{cen_quasar_2000,madau_earliest_2000,haiman_probing_2001,fan_constraining_2006}}. 

Not only does the proximity effect lead to a degeneracy in neutral gas fraction constraints, the observed proximity zones also contain a wealth of information about the quasar's past activity. Proximity zone sizes have proved particularly useful in inferring the time period over which a given quasar has been UV luminous, known as the quasar lifetime. Measurements of quasar lifetimes from single-epoch spectra of high-redshift quasars have yielded an average quasar lifetime of $\log_{10}(t_{\rm Q}/\rm yr) = 5.7_{-0.3}^{+0.5}$ 
for a sample of 15 quasars at $5.8 \leq z \leq 6.6$ \citep{morey_estimating_2021}, which, along with other lifetime measurements of individual objects \citep{eilers_first_2018,eilers_detecting_2021,davies_evidence_2019,davies_constraining_2020,andika_probing_2020}, has challenged our models of supermassive black hole (SMBH) growth. These measurements have also revealed a population of surprisingly young quasars with billion-solar-mass black holes at $z\gtrsim 6$ \citep{eilers_detecting_2021}, which pose an even greater strain on our models of black hole growth, suggesting that either super-Eddington accretion or UV-obscured growth phases need to be evoked \citep{eilers_first_2018,davies_evidence_2019,satyavolu_need_2023}.
This imminent challenge calls for further investigation into population-level quasar lifetimes that is symbiotic with constraining the progress of reionization.

To this end, we use a sample of 18 spectra of quasars from the Epoch of Reionization 
to constrain the evolution of the neutral hydrogen gas fraction, $\vxhi$, at redshifts of $6.0 < z < 7.1$. We overcome the challenges of \change{insensitivity and patchiness of reionization} by stacking our quasar sample in discrete redshift intervals \change{in order to increase the signal-to-noise of the average damping wing signature in our sample. Therefore, we fit} an average quasar spectrum in contrast to existing studies in the literature which focus on fitting single-sightline spectra. Additionally, we take into account the degeneracy between the neutral gas fraction and quasar radiation imprints by jointly fitting for a \change{sample-}average quasar lifetime in each redshift bin. 
Our study showcases the emergence of damping wings near the end of the EoR and thus presents an important step towards measuring sightline-averaged constraints of the volume-averaged neutral fraction, as well as average quasar lifetimes across cosmic time. 

This paper is structured as follows. In \S~\ref{sec:methods}, 
we describe our observations, as well as the reconstruction method of the intrinsic emission around the \lya line of our observed quasars, in order to obtain normalized flux transmission profiles. In \S~\ref{sec:sims}, we describe the simulations that we use to model our observations, and we provide the details about their inference in \S~\ref{sec:inference}. Subsequently, we present the neutral fraction and quasar lifetime constraints in \S~\ref{sec:results}, and provide a short summary in \S~\ref{sec:summary}.

\begin{figure*}[h!]
    \centering
    \includegraphics[width=0.9\linewidth]{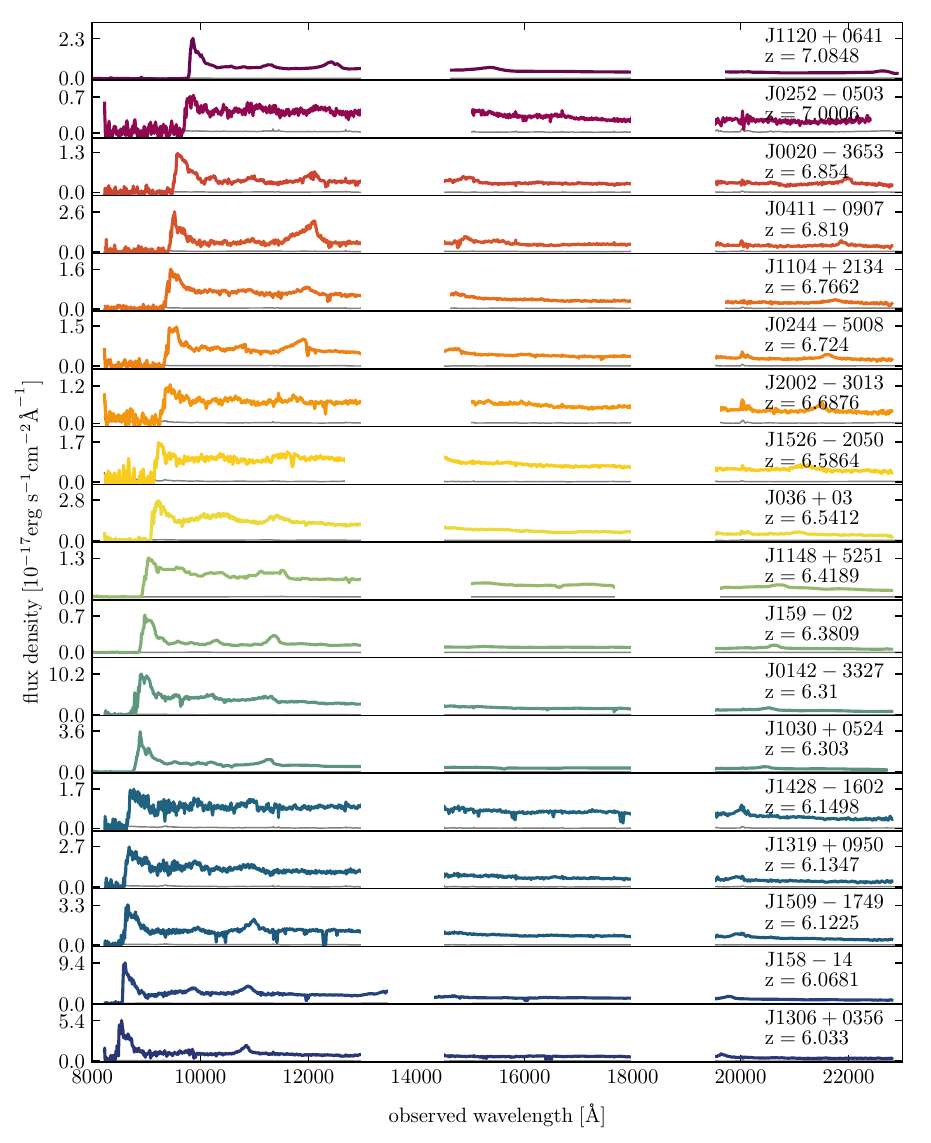}
    \caption{The 18 quasar spectra used in this study sorted and color-coded by redshift. The error vector is shown as a thin gray line. }
    \label{fig:FIRE_composite}
\end{figure*}

Throughout this paper, we use the flat $\Lambda$CDM cosmology with $h = 0.67$, $\Omega_M=0.31$, $\Omega_\Lambda=0.69$ \citep{planck_collaboration_planck_2020}.

\section{Data}\label{sec:methods}

\subsection{Quasar sample}

This study uses a sample of 18 quasars at redshifts $6.0 < z < 7.1$, shown in \cref{fig:FIRE_composite}. The spectra for the majority of our target quasars were taken with the Magellan/FIRE spectrograph \citep{simcoe_fire_2013}. For three of the quasars in our sample, i.e. J1030+0524, J159--02, and J1120+0641, we use combined Magellan/FIRE and VLT/X-Shooter \citep{vernet_x-shooter_2011} spectra, while one other spectrum in our sample, i.e. J1148+5251, was observed with Keck/MOSFIRE \citep{mclean_design_2010,mclean_mosfire_2012} and Keck/ESI \citep{sheinis_esi_2002}. The total integration time along with other information and references is included in \cref{tab:qso_table}. Note that the highest-redshift quasar in our sample is ULAS J1120+0641, which has been studied extensively \citep[e.g.][]{simcoe_extremely_2012,bosman_deep_2017,schindler_x-shooteralma_2020} as it was the first $z>7$ quasar ever discovered \citep{mortlock_luminous_2011}. The spectrum of ULAS J1120+0641 that we are using in this paper has been observed with Magellan/FIRE and VLT/X-Shooter, as compared to the VLT/FORS and Gemini/GNIRS spectrum that has been originally published for this object.

All spectroscopic data were reduced using the \texttt{PypeIt} package\footnote{\url{https://pypeit.readthedocs.io/en/latest/}} \citep{pypeit:joss_pub, pypeit:zenodo}. In particular, we subtract flat fields that were taken during the same observing run and perform sky subtraction by differencing dithered A-B exposure pairs with a further sky line residual elimination according to \citet{bochanski_mase_2009}. The subtracted 2D images then undergo optimal extraction \citep{horne_optimal_1986} to obtain the 1D spectra. The wavelength solution is derived from the night sky OH lines. Subsequently, all extracted 1D spectra are flux calibrated by using sensitivity functions from standard stars. For each quasar, we coadd the flux-calibrated 1D spectra across exposures and correct for telluric absorption by jointly fitting an atmospheric model and a quasar model.

A bias in the inferred neutral gas fraction occurs if sightlines to the quasars in our sample intercept dense neutral hydrogen clouds, called proximate damped \lya systems (pDLAs). Such high neutral hydrogen column density can cause damping wings to appear even in a fully ionized Universe, thus mimicking the effect of a highly neutral IGM. \change{All quasar spectra in this sample were checked for metal lines that would indicate a presence of a proximate absorber} -- for example, J183+05 had to be removed from our sample \citep{banados_metal-poor_2019}. 

\begin{figure}
    \centering
    \includegraphics{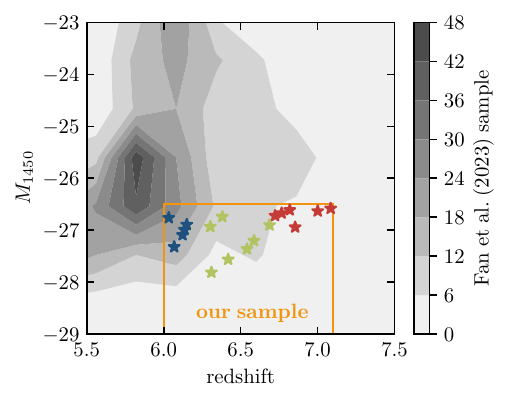}
    \caption{The magnitude vs.\ redshift distribution of quasars in our sample (colors corresponding to the three different redshift bins used in our analysis) compared to the high-redshift quasar database published by \cite{fan_quasars_2023} shown as gray contours.}
    \label{fig:M1450vsz}
\end{figure}

\change{In addition to the neutral gas fraction and the quasar lifetime, the luminosity of the quasar also affects the quasar transmission profiles -- a larger ionizing photon output leads to a larger proximity zone in a shorter time. While we forward model the luminosities in our quasar sample (see \S~\ref{sec:inference}), we imposed a luminosity cut of $M_{1450}=-26.5$ to ensure that the individual transmission profiles are more directly comparable. This luminosity cut also allowed us to obtain good quality spectra in the limited observing time available.} The distribution of our quasars in the magnitude--redshift space is shown in \cref{fig:M1450vsz}, along with a comparison of our sample to a recently published sample \citep{fan_quasars_2023}. Quasars in our sample are color-coded to show the three discrete redshift bins, $6.0 \leq z < 6.3$ (blue), $6.3 \leq z < 6.7$ (green), and $6.7 \leq z \leq 7.1$ (red), that correspond to our final constraints.

\begin{table*}
    \centering
    \caption{Quasar sample used in this study. \change{For each object, we include its coordinates, the total observed time, its redshift ($z$) and redshift uncertainty ($\sigma_z$), its absolute magnitude at rest-frame $1450\ \mathrm{\AA}$ ($M_{1450}$), its SNR, as well as references to its discovery paper and its redshift measurement (including the emission line used to determine the quasar's redshift).} F01 - \cite{fan_survey_2001}, F03 - \cite{fan_survey_2003}, W03 - \cite{white_probing_2003}, M05 - \cite{maiolino_first_2005}, K07 - \cite{kurk_black_2007}, W07 - \cite{willott_four_2007}, M09 - \cite{mortlock_discovery_2009}, M11 - \cite{mortlock_luminous_2011}, B15 - \cite{banados_bright_2015}, C15 - \cite{carnall_two_2015}, V15 - \cite{venemans_identification_2015}, B16 - \cite{banados_pan-starrs1_2016}, M17 - \cite{mazzucchelli_physical_2017}, C18 - \cite{chehade_two_2018}, D18 - \cite{decarli_alma_2018}, R19 - \cite{reed_three_2019}, P19 - \cite{pons_new_2019}, W19 - \cite{wang_exploring_2019}, Y19 - \cite{yang_exploring_2019}, E21 - \cite{eilers_detecting_2021}, V20 - \cite{venemans_kiloparsec-scale_2020}, Y20 - \cite{yang_measurements_2020}, Y21 - \cite{yang_probing_2021}, M23 - \cite{marshall_black_2023}. Note that J1030+0524, J159--02, and J1120+0641 are combined Magellan/FIRE and VLT/X-shooter spectra, and J1148+5251 is a Keck/MOSFIRE and Keck/ESI spectrum.}
    \label{tab:qso_table}
    \centering
    \begin{tabular}{l l l l l l l l l l l}
        \hline
        Quasar & R.A. & Dec. & Obs. time & $z$ & $\sigma_z$ & $M_{1450}$ & \change{SNR\footnote{This is the mean SNR of the continuum per a $100$km/s resolution element in the wavelength range $1270 \mathrm{\AA} - 1290 \mathrm{\AA}$.}} & Discovery ref. & $z$ ref. & $z$ note \\
         & J2000.0 & J2000.0 & [h] & & & & & \\ \hline
        %J0410--0139 & 00:00:00.000 & --00:00:00.000\footnote{Quasar will be published in Ba\~nados et al.\ (in prep).} & $5.8$ & $6.99$ & $0.05$ & $-28.12$ & Bp & Bp & \lya \\ 
        J0020--3653 & 00:20:31.470 & --36:53:41.800 & $8.0$ & $6.854$ & $0.002$ & $-26.94$ & $3.9$ & R19 & M23 & \oiii \\ 
        J0142--3327 & 01:42:43.700 & --33:27:45.720 & $3.3$ & $6.31$ & $0.03$ & $-27.81$ & $5.3$ & C15 & C15 & \lya \\ 
        J036+03 & 02:26:01.873 & +03:02:59.254 & 2.0 & $6.5412$ & $0.0018$ & $-27.36$ & $5.0$ & V15 & B15 & \cii \\ 
        J0244--5008 & 02:44:01.020 & --50:08:53.700 & $8.8$ & $6.724$ & $0.002$ & $-26.72$ & $7.2$ & R19 & R19 & \mgii \\ 
        J0252--0503 & 02:52:16.640 & --05:03:31.800 & $2.2$ & $7.0006$ & $0.0009$ & $-26.63$ & $2.7$ & Y19 & Y21 & \cii \\ 
        J0411--0907 & 04:11:28.630 & --09:07:49.800 & $4.3$ & $6.819$ & $0.002$ & $-26.61$ & $2.8$ & W19, P19 & M23 & \oiii \\ 
        J1030+0524 & 10:30:27.098 & +05:24:55.000 & $9.2$ & $6.303$ & $0.001$ & $-26.93$ & $37.4$ & F01 & D18 & \mgii \\ 
        J158--14 & 10:34:46.509 & --14:25:15.890 & $2.7$ & $6.0681$ & $0.0001$ & $-27.32$ & $3.1$ & C18 & E21 & \cii \\ 
        J159--02 & 10:36:54.191 & --02:32:37.940 & $8.2$ & $6.3809$ & $0.0005$ & $-26.74$ & $14.6$ & B16 & D18 & \cii \\ 
        J1104+2134 & 11:04:21.590 & +21:34:28.800 & $8.5$ & $6.7662$ & $0.0009$ & $-26.67$ & $6.0$ & W19 & Y21 & \cii \\ 
        J1120+0641 & 11:20:01.479 & +06:41:24.300 & $33.0$ & $7.0848$ & $0.0004$ & $-26.58$ & $22.4$ & M11 & V20 & \cii \\ 
        J1148+5251 & 11:48:16.652 & +52:51:50.440 & $26.2$ & $6.4189$ & $0.0006$ & $-27.56$ & $72.0$ & F03, W03 & M05 & \cii \\ 
        J1306+0356 & 13:06:08.260 & +03:56:26.300 & $4.7$ & $6.0330$ & $0.0002$ & $-26.76$ & $2.7$ & K07 & D18 & \cii \\
        J1319+0950 & 13:19:11.300 & +09:50:51.490 & $5.3$ & $6.1347$ & $0.0005$ & $-26.99$ & $3.2$ & M09 & V20 & \cii \\
        J1428-1602 & 14:28:21.390 & -16:02:43.300 & $4.3$ & $6.1498$ & $0.0011$ & $-26.89$ & $2.6$ & B16 & D18 & \cii \\
        J1509-1749 & 15:09:41.780 & -17:49:26.800 & $5.0$ & $6.1225$ & $0.0007$ & $-27.09$ & $3.2$ & W07 & D18 & \cii \\
        J1526--2050 & 15:26:37.840 & --20:50:00.700 & $6.5$ & $6.5864$ & $0.0005$ & $-27.20$ & $3.6$ & M17 & D18 & \cii \\ 
        J2002--3013 & 20:02:41.590 & --30:13:21.700 & $5.3$ & $6.6876$ & $0.0004$ & $-26.90$ & $4.2$ & Y20 & Y21 & \cii \\ \hline
    \end{tabular}
\end{table*}

\begin{figure*}[h!]
    \centering
    \includegraphics[width=\linewidth]{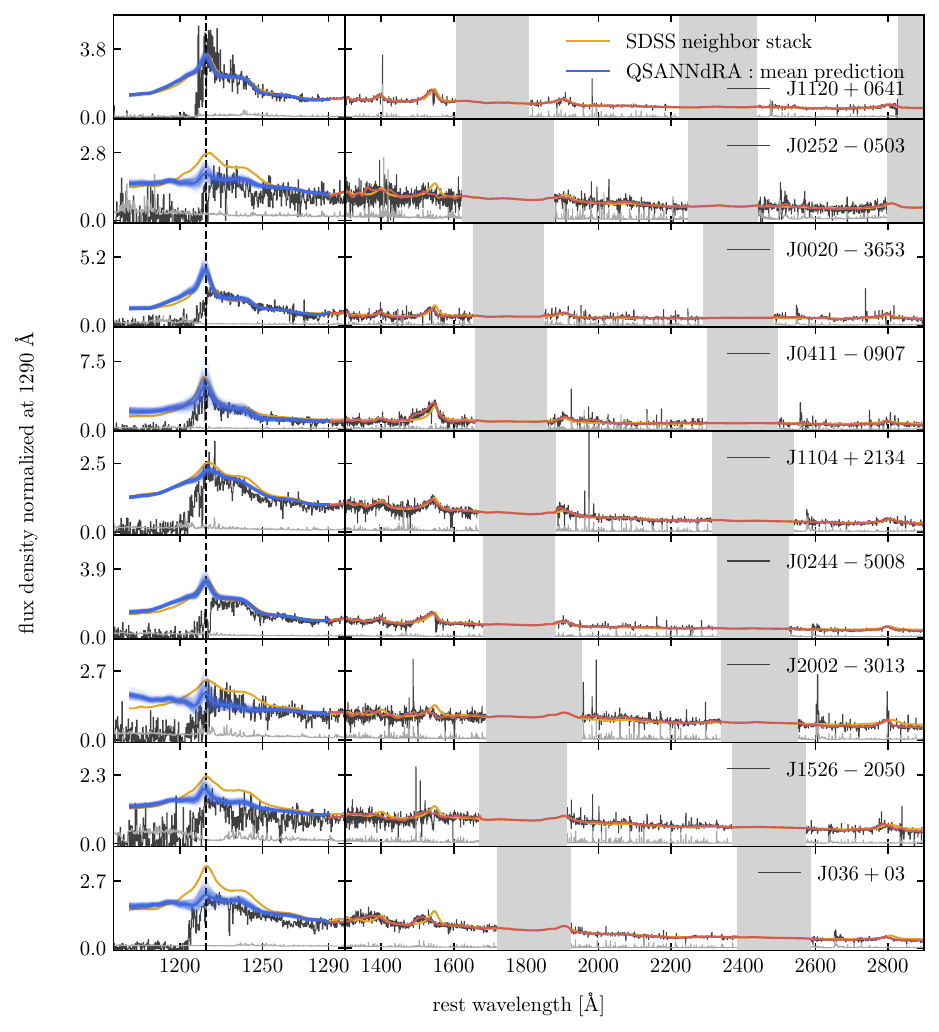}
    \caption{Quasar continuum reconstruction for all objects in our sample predicted by QSANNdRA. We show the observed data in black and the measurement noise vector in light gray. QSANNdRA outputs a range of predictions (in light blue) as well as a mean prediction (dark blue) based on the red spectral fit above $1290$ \AA\ (red curves). The flux in the shaded telluric regions is determined from the nearest neighbour SDSS stack shown in orange.}
    \label{fig:QSANNdRA_predictions}
\end{figure*}

\begin{figure*}\ContinuedFloat
    \centering
    \includegraphics[width=\linewidth]{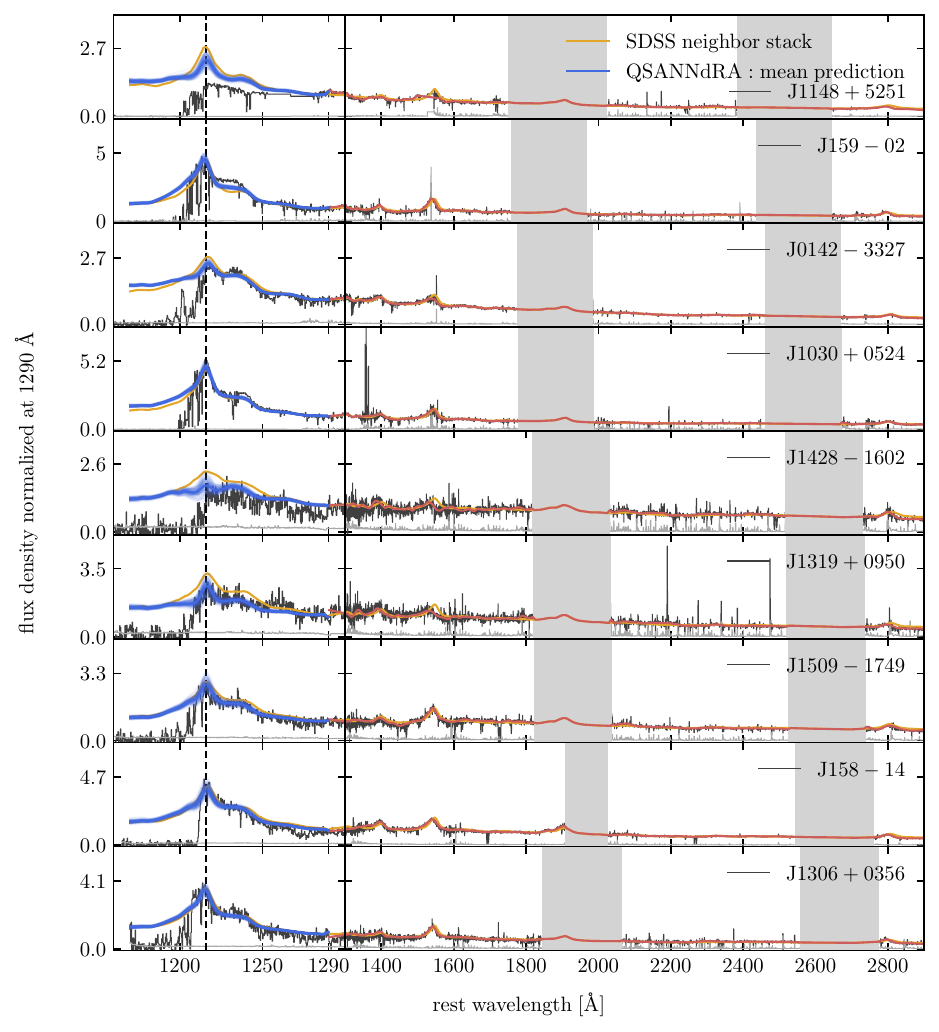}
    \caption{(contd.)}
\end{figure*}

\subsection{\lya continuum reconstruction}\label{sec:cont}

In order to constrain the volume-averaged neutral fraction and the quasar lifetime, we need to first estimate the intrinsic quasar continua around the \lya line. To this end, we implement the neural network model called QSANNdRA, which was first introduced by \cite{durovcikova_reionization_2020}. Around \lya, this model has been shown to produce a minimal bias, $0.3\%$, and a scatter below 10\% at $1\sigma$, competitive with the best continuum reconstruction techniques in the literature \citep{bosman_comparison_2021}.

In this work, we use QSANNdRA to predict the intrinsic quasar continuum in the wavelength range $1170 \mathrm{\AA} - 1290 \mathrm{\AA}$ (the ``blue side'') based on the ``red-side'' spectrum at $1290 \mathrm{\AA} - 2900 \mathrm{\AA}$, which is agnostic to absorption by the neutral hydrogen in the intervening IGM along the line-of-sight.

We train and test this algorithm on low-redshift quasar spectra from the sixteenth data release version of the SDSS Quasar Catalog \citep{lyke_sloan_2020}, and perform the same set of training data cuts, the same spectral smoothing procedure as well as the same network architecture and training hyperparameters as in the original work \citep{durovcikova_reionization_2020}. In summary, we utilize the quasar catalog flags to exclude all broad-absorption line (BAL) quasars and quasars with known proximate damped Lyman absorbers (pDLAs) in their sightlines. \change{Additionally, we impose a signal-to-noise cut of $\geq 3.0$ using the \texttt{SN\_MEDIAN\_ALL} flag that represents the median flux density-to-noise across all good pixels in the spectrum \citep{lyke_sloan_2020}}. We choose the low-redshift spectra to have redshifts of $2.09 < z < 2.51$ such that both the \lya and the \ion{Mg}{2} emission lines fall into the spectral range of the BOSS and SDSS spectrographs. In order to recover smooth spectral continua without absorption lines, we smooth each spectrum with an updated version of the QSmooth algorithm\footnote{\url{https://github.com/DominikaDu/QSmooth}} \citep{durovcikova_reionization_2020} that implements piecewise spline fitting in the last smoothing step. We further clean up our training set with the Random Forest \citep{breiman_random_2001} rejection method described in \cite{durovcikova_reionization_2020}. The aforementioned steps are implemented using the Scikit-Learn Python package \citep{pedregosa_scikit-learn_2011}.

The remaining training set consists of 28,443 quasars, of which $80\%$ are used for training and the remaining $20\%$ are used for testing. QSANNdRA is constructed and trained using the Keras Python package \citep{chollet2015keras}. The trained model achieves a mean absolute error of $\lesssim 14\%$ across the whole prediction wavelength range. Moreover, we checked that this performance is nearly identical for the train and test data sets, which suggests good generalizability of this model to previously unseen spectroscopic data.

The trained model is subsequently applied to our high-redshift quasar sample. First, we rebin the high-$z$ spectra to a $100\mathrm{km/s}$ resolution, which roughly matches the spectral resolution of the low-redshift SDSS sample, and we smooth each spectrum using QSmooth as described earlier. Since their rest-frame ultraviolet spectra are redshifted to the near-infrared, our high-$z$ spectra all contain regions of enhanced telluric absorption due to the Earth's atmosphere which we mask before smoothing. Specifically, we mask out the following regions: $13,000 \mathrm{\ \AA} \lesssim \lambda_\mathrm{obs} \lesssim 14,500 \mathrm{\ \AA}$, $18,000 \mathrm{\ \AA} \lesssim \lambda_\mathrm{obs} \lesssim 19,500 \mathrm{\ \AA}$, and $ \lambda_\mathrm{obs} \gtrsim 22,800 \mathrm{\ \AA}$. We visually inspected each smoothed red-side spectrum, and for a number of quasars we manually adjusted smoothing parameters, the telluric mask, and masked additional absorption features that seemed to be biasing the QSmooth continuum.

Due to QSANNdRA's inability to work with missing input data, we fill in the telluric regions in each quasar spectrum with a mean spectral stack of its $100$ nearest neighbours in the SDSS training sample. To identify these nearest neighbours, we perform a simple Nearest Neighbor search using the Scikit-Learn package \citep{pedregosa_scikit-learn_2011} in a compressed, 10-component PCA space of their red-side spectrum. \change{In order to ensure a continuous fit at the edges of the telluric mask, we perform a separate spline smoothing in a narrow window around each edge.}

After filling in the telluric gaps, we apply QSANNdRA to all high-$z$ quasars in our sample and display the resultant continuum predictions in \cref{fig:QSANNdRA_predictions}. Based on the input red-side spectrum (shown in red), the model outputs $100$ individual samples of the continuum (shown in faint blue), as well as the mean prediction for each quasar (thick blue). The SDSS nearest-neighbor stack is shown in orange for reference, but is not used in the remainder of this work. Note that differences between the SDSS composites and the predictions from QSANNdRA (and other continuum reconstruction techniques) have been seen before (e.g. \citet{bosman_comparison_2021}) and are not surprising -- \change{in fact, the existence of these discrepancies motivates the existence of more sophisticated continuum prediction models. Note that} the biggest differences in the \lya\ continuum are seen for quasars whose \civ\ line at $\lambda_{\rm rest} = 1545.86\ \mathrm{\AA}$ does not match the SDSS neighbor stack very well\change{, showcasing a limitation of the low-redshift SDSS training set}. This mismatch is thus a manifestation of the mild redshift evolution in the \civ\ line properties, as has been noted in \citet{fan_quasars_2023} (see also \citet{meyer_new_2019,schindler_x-shooteralma_2020}). Also note that we ultimately forward model the redshift uncertainty in our later analysis and thus the exact redshift that is used to bring the quasars into their rest frame should not affect our results.

We discuss and display these continuum predictions in more detail in Appendix \ref{app:continua_zoom}.

\subsection{Stacking Quasar Spectra}

With the continuum predictions at hand, we proceed by dividing our quasar sample into discrete redshift intervals and stacking all continuum-normalized spectra within each bin to compute an average transmitted quasar spectrum. \change{We choose to stack our transmission profiles for two main reasons, as opposed to analysing each quasar sightline individually: First, averaging multiple quasar spectra allows us to capture the potentially weak average damping wing signature in the presence of uncertainties on the continuum prediction. While individual $\vxhi$ constraints would only be possible for quasars with exceptionally strong damping wings (for which we happen to probe a highly neutral line of sight), our method allows us to use information across multiple sightlines sampling a range of column densities to constrain the volume-averaged neutral gas fraction within a given redshift range. Secondly, quasar environments and the topology of reionization are not Gaussian processes, which complicates constraining the neutral gas fraction. Single-sightline studies tackle this problem by rebinning the quasar transmission profiles to a lower resolution (i.e. coarse-graining) in order to randomize the noise in each pixel. Averaging over multiple of such coarse-grained profiles (i.e. stacking) thus helps us decrease the noise in the average damping wing signature and further justifies its approximation as a multivariate Gaussian distribution during inference (see \S~\ref{sec:inference}).} 

% Stacking has two main advantages: Firstly, it allows us to account for the large cosmic variance near the end of reionization and thus overcome the insensitivity to the intermediate range of neutral gas fractions, $\vxhi$. Note that all damping wing based neutral fraction constraints to date \citep{wang_luminous_2021,greig_ly_2017,greig_constraints_2019,durovcikova_reionization_2020,davies_quantitative_2018,banados_800-million-solar-mass_2018,yang_poniuaena_2020} only sample single sightlines in the sky, which impedes their usability near the end of the EoR when the neutral gas fraction is very patchy. Secondly, stacking and averaging over multiple density fields helps with ``gaussianizing'' the noise in the spectrum, which allows us to approximate the likelihood function used for inference as a multivariate Gaussian distribution (see \S~\ref{sec:inference}). 

Due to the aforementioned degeneracy between the neutral fraction and quasar radiation effects, we also obtain a sample-average constraint on the quasars' lifetimes. This stacking procedure is inspired by \cite{morey_estimating_2021}, who previously used stacking to determine the mean quasar lifetime in a sample of 15 quasars at $5.8 \leq z \leq 6.6$. Note that the interpretation of the inferred mean quasar lifetimes depends on the state of the IGM; for a fully ionized IGM, proximity zone sizes are sensitive to the episodic lifetime (i.e. the most recent ``on'' phase), whereas a mostly neutral IGM enables a measurement of the integrated quasar lifetime \citep{eilers_implications_2017}. Therefore, due to stacking, we expect to be more sensitive to the episodic lifetime around $z\sim 6$, and the integrated lifetime towards $z\sim 7$, although the exact interpretation is complicated. \change{Another factor that affects the interpretation of quasar lifetimes is the geometry of the quasar emission relative to the observer's line of sight, as the proximity zone size need not be isotropic. We stress that our method is only sensitive to the line-of-sight proximity zone size, and so we will henceforth assume an isotropic proximity zone for the interpretation of our results.}

\begin{figure}
    \centering
    \includegraphics{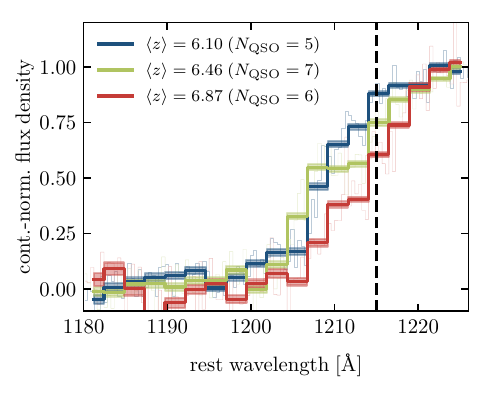}
    \caption{Stacked spectra of quasars in our sample in three distinct redshift bins show the emergence of damping wings between $z\sim 6$ and $z\sim 7$ (thick colored curves, the location of \lya is marked by the dashed black line). The number of quasars within a given redshift bin is noted in brackets, and the shaded region around the stacked spectra shows the variance in the mean flux. The thin colored lines show a higher-resolution ($\Delta v = 100\ {\rm km/s}$) version of the three stacked spectra.}
    \label{fig:zstacks}
\end{figure}

The resultant stacked spectra are shown in \cref{fig:zstacks}. We show spectral stacks in three redshift bins, $6.0 \leq z < 6.3$ (blue), $6.3 \leq z < 6.7$ (green), and $6.7 \leq z \leq 7.1$ (red). These redshift intervals were chosen such that each bin contains approximately the same number of objects while avoiding bins of vastly unequal width. This is particularly relevant at the lower end of our redshift range, where we expect damping wings to have not yet fully emerged. Crucially, we use histogram binning with a resolution of $600\ {\rm km/s}$ to avoid correlating the uncertainties in neighbouring pixels and to smooth over the density fields in the vicinity of the quasars. \change{Note that the purpose here is to constrain the average neutral gas fraction in each redshift bin, while avoiding making any statements about the evolution of $\vxhi$ within each redshift interval.}

It should be noted that our analysis is still limited to a small sample of quasars along different sightlines. However, by performing this stacking procedure we highlight the possibility of using damping wings of quasars also near the end of the EoR.

\section{Hydrodynamical Simulations}\label{sec:sims}

In order to model both the effect of the neutral gas in the surrounding IGM and the effect of the quasar radiation field, we employ hydrodynamical simulations with radiative transfer. We use two different reionization simulations, namely the P-GADGET-3 simulations post-processed with the ATON code for radiative transfer (henceforth referred to as ATON, \citet{satyavolu_need_2023}), and the CROC simulations with fully coupled Optically Thin Variable Eddington Tensor (OTVET) radiative transfer (henceforth referred to as CROC, \citet{chen_distribution_2021}). In both cases, we insert a mock quasar into the most massive halos in the simulation snapshots and draw multiple sightlines in order to simulate cosmic variance\footnote{\change{Note that we use the term ``cosmic variance'' to denote the variance in sampling different lines of sight and hence different density environments of quasars both in the sky and in the simulations.}} of the line-of-sight density and ionization fields. The following sections provide further details about the two simulations and a brief discussion of their main differences. Our use of two simulations is motivated by checking for consistency and increasing the robustness of our results, and we stress that the aim of this paper is not to assess the validity of either of them.

\subsection{P-GADGET-3 simulations with ATON radiative transfer}

The first suite of models comes from post-processing the P-GADGET-3 simulation (modified version of GADGET-2, \citet{springel_cosmological_2005}) using the ATON code \citep{aubert_radiative_2008,aubert_reionization_2010} for three-dimensional radiative transfer \citep{kulkarni_large_2019}. ATON solves the radiative transfer using the M1 approximation \citep{aubert_radiative_2008,gnedin_multi-dimensional_2001} and obtains the gas ionized fraction and temperature self-consistently. All our ionizing sources are placed in halos of mass $\gtrsim 10^9\ $\msol. The radiative transfer is run with a single photon frequency to reduce computational cost. The box size of this simulation is $160\ \chimp$ with $2048^3$ gas and dark matter particles and a resolution of $78.125\ {\rm ckpc}/h$. The simulation is run between $z=99$ and $z=4$, and physical quantities such as the gas densities are saved in intervals of $40\ {\rm Myr}$. As discussed in \citet{kulkarni_large_2019} and \citet{keating_long_2020}, these simulations are calibrated to match the observed mean \lya flux at $z>5$ \citep{bosman_new_2018,becker_evidence_2015} and are consistent with a number of high-redshift observations \citep{planck_collaboration_planck_2020,keating_long_2020,becker_evidence_2015,greig_ly_2017,greig_constraints_2019,davies_quantitative_2018,wang_significantly_2020,weinberger_lyman-_2018,weinberger_modelling_2019,gaikwad_probing_2020}. The end-point of reionization in this simulation is at $z=5.3$, with the process half complete at $z=7$. This picture remains consistent with the latest observations of the effective \lya opacity by \cite{bosman_hydrogen_2022}. 

The creation of synthetic quasar spectra is described in \cite{satyavolu_need_2023}. To save computational costs, we only use the simulation snapshots with neutral fractions of $\vxhi = \{3.7\times 10^{-5},$ $0.07,$ $0.13,$ $0.21,$ $0.37,$ $0.54,$ $0.75,$ $0.89 \}$ in this work. In each snapshot, quasars are placed in halos having masses in the range $10^{11}$~\msol~$\lesssim M_{\mathrm{halo}} \lesssim 10^{12}$~\msol \footnote{There are $>7000$ halos in this mass range at simulation redshift of $z_{\rm sim} = 5.95$.}. \change{The same set of halos are utilized for all quasars in our sample.} The quasar light curve is assumed to be a `light bulb'. For each of the quasars in our sample, the magnitude at $1450\ {\rm \AA}$ ($M_{1450}$) is converted into the total number of ionizing photons $\dot{N}_{\rm tot}$ by assuming the quasar to have a broken power-law spectral profile, i.e.
\begin{equation}\label{eq:Ntot}
	\dot{N}_{\rm tot}=\int^{\infty}_{\rm 13.6 eV}\frac{{L}_{\nu}}{h\nu}d\nu \quad ; L_\nu\propto \nu^{\alpha_\nu},
\end{equation}
\change{with the spectral index $\alpha_\nu=-0.61$ and a break at $\lambda = 912\ \mathrm{\AA}$ corresponding to the hydrogen ionization energy \citep{lusso_first_2015}.}
The simulated quasar lifetimes have been chosen to have the following values: $\logtq = \{4.0,5.0,6.0,7.0,8.0\}\ {\rm Myr}$.

For each quasar in our sample, we rescale the physical length scales in each simulation snapshot by a factor of $(1+z)/(1+z_{\rm sim})$ and the densities along the line of sight by a factor of $(1+z)^3/(1+z_{\rm sim})^3$, where $z$ is the quasar redshift and $z_{\rm sim}$ is the redshift of the simulation snapshot. We then perform a line of sight radiative transfer using the method described in \cite{satyavolu_need_2023} and draw between $500$ and $1000$ sightlines in each snapshot. This method involves solving the thermochemistry equations to determine the abundances of ionized hydrogen and helium, alongside tracking the evolution of temperature. We use the post-processed neutral gas fraction, temperature, along with the rescaled gas densities and peculiar velocities from the underlying cosmological simulation to compute the \lya optical depth by assuming a Voigt profile \citep{garcia_voigt_2006}. 

\subsection{CROC simulations with OTVET radiative transfer}

The second suite of models is generated by post-processing the \textit{Cosmic Reionization on Computers} (CROC) simulations \citep{gnedin_cosmic_2014}.
We utilize all six $40\ \chimp$ CROC simulation runs to sample a wide range of large-scale structures. All these simulations are run with the same physics with the sole difference being the random seed utilized to generate the initial conditions. Therefore, they should be treated as six typical random places in the universe. These CROC simulations are run with the Adaptive Refinement Tree (ART) code \citep{kravtsov_high-resolution_1999, kravtsov_constrained_2002, rudd_effects_2008} to reach high spatial resolution using the adaptive mesh refinement approach. The base grid is $39\ {\rm ckpc/h}$ in size, and the peak resolution is $\approx 100\ {\rm pc}$ (in physical units). CROC simulations include relevant physics such as gas cooling, heating, star formation and stellar feedback. In the simulations, individual star particles are the main radiation sources and the radiative transfer is done using the Optically Thin Variable Eddington Tensor (OTVET) method \citep{gnedin_multi-dimensional_2001}, which is fully coupled to gas dynamics. 

To create synthetic spectra for quasars, we post-process the sightlines with 1D radiative transfer code described in \citet{chen_distribution_2021}. The sightlines are drawn from the full simulation snapshots. The full snapshots contain complete information of the simulation such as neutral fraction of hydrogen and helium, density, temperature and peculiar velocity of the gas. Because of the limited storage, full simulation snapshots are saved sparsely, and thus we only have 4 full snapshots with volume-averaged neutral fraction $\vxhi$ between $0.99$ and $10^{-4}$ for each run. Because of this, we use all six boxes to create a more regular grid of $\vxhi$ by binning the snapshots into groups around the following values: $\vxhi = \{4\times 10^{-4},$ $0.10,$ $0.36,$ $0.57,$ $0.83,$ $0.94,$ $0.98 \}$, such that each $\vxhi$ bin contains $2$-$3$ snapshots. In addition to making the neutral fraction grid more regular, we do this to increase the cosmic variance captured by the CROC models at a given neutral gas fraction, while keeping the variance comparable across the different neutral fraction models.

From each simulation snapshot \change{and each quasar}, we select the $20$ most massive halos and draw $50$ random sightlines centered on them. The masses of these halos at $z=6.8$ are between $1.1\times 10^{11} M_\odot <M_h<1.1\times 10^{12} M_\odot$.
To simulate each quasar, we keep the neutral fraction and temperature of each cell unchanged while we scale the physical length and density to the redshift of the quasar by the expansion factor (same as in the ATON simulation). The total number of ionizing photons is calculated from $M_{1450}$ by adopting the average broken power-law spectral shape measured by \citet{lusso_first_2015}, see \cref{eq:Ntot} (again, same as in ATON). We post-process the sightlines in each snapshot for the following range of quasar lifetimes: $\logtq = \{3.0,$ $3.5,$ $4.0,$ $4.5,$ $5.0,$ $5.5,$ $6.0,$ $6.5,$ $7.0,$ $7.5,$ $7.8,$ $8.0\}\ {\rm Myr}$. Note that for the highest-$z$ bin, all the $50$ sightlines per halo are post-processed with quasar parameters according to the observed values, however, due to computational limitations only $5$ sightlines per halo are post-processed for most of the quasars in the two lower redshift bins. We checked that changing the exact number of sightlines does not significantly alter our results.

\begin{figure*}[t!]
    \centering
    \includegraphics[width=\linewidth]{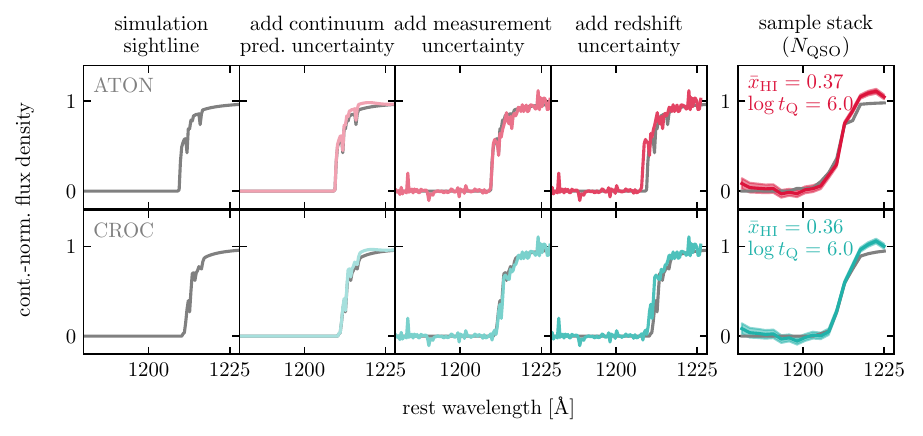}
    \caption{Illustration of the forward modeling of uncertainties and model construction for both simulations (ATON in the top panel, CROC in the bottom panel). For a particular quasar and a combination of $\vxhi$ and $\logtq$ values, we draw one sightline at random from the corresponding simulation (first column from the left). We forward model the uncertainties by adding the smooth continuum prediction \change{uncertainty} (second column), random measurement \change{uncertainty} (third column), and finally the corresponding redshift \change{uncertainty} along the x-axis (fourth column; $\sigma_z = 0.01$ for illustration purposes). We repeat this sightline processing for all quasars within a given redshift bin ($N_{\rm QSO}$), and then stack $N_{\rm QSO}$ processed sightlines to form a sample stack (rightmost column\change{; here we also show the sample stack without any added uncertainties in gray}). The whole procedure is repeated $10000$ times for the same combination of $\vxhi$ and $\logtq$ in order to calculate the final model and noise statistics that are used for inference.}
    \label{fig:noise}
\end{figure*}

\subsection{Comparison of CROC and ATON}\label{sec:comp}

This section summarizes the main similarities and differences of the two hydrodynamical simulations employed in this work. We emphasize that the aim of this paper is not to perform an in-depth comparison of reionization simulations, but rather to use the two simulations to estimate systematic uncertainties and increase the robustness of our constraints.

In both simulations, we choose the most massive halos to host mock quasars, each of which is magnitude and redshift matched to quasars in our sample. Note that redshift-matching amounts to rescaling the physical scales and the density fields to the redshift of the quasar across snapshots at different simulation redshifts -- the simulation redshift is only a proxy to the volume-averaged neutral gas fraction. \change{Note, however, that the population of ionizing sources may vary across simulation redshifts between the two simulations, which can result in diverse reionization morphologies \citep{cain_morphology_2023} and can affect the sightline properties. Employing two different simulations therefore helps us increase the robustness of our results in the presence of systematics such as these.} Furthermore, in both simulations we use the same broken power law spectral template \citep{lusso_first_2015} to convert the magnitudes of the quasars in our sample to the number of ionizing photons.

The main differences between the two simulations are how radiative transfer is treated, the simulated volumes, and the resolution. The CROC simulation implements a 3D radiative transfer that is fully coupled to gas dynamics, whereas in ATON the radiative transfer is only done in postprocessing. \citet{satyavolu_need_2023} tested how using their 1D radiative transfer impacts the hydrogen and helium ionization fractions and the gas temperature as compared to a full 3D radiative transfer, and found the difference minimal.

The size of the simulated box is four times larger in ATON ($160\ \chimp$) than CROC ($40\ \chimp$), which can have an impact on the sizes of ionized bubbles that are modelled in each case. \citet{iliev_simulating_2014} has shown that ionized bubbles require box sizes of $>100\ \chimp$ to converge. Smaller simulation volumes can thus contain systematically smaller ionized bubbles and hence produce stronger damping wings in the mock sightlines \citep[see][]{keating_jwst_2023}.

Lastly, the high-spatial resolution of the CROC simulation produces DLAs which can also bias the simulated sightlines towards stronger damping wings. To reduce this bias, we exclude all simulated sightlines with neutral hydrogen column densities of $N_{\rm HI} > 5\times 10^{19}\ {\rm cm}^{-2}$ (motivated by \citet{chen_distribution_2021}). Note that we also checked the impact of an increased spatial resolution on the resultant sightlines in the lower-resolution ATON simulation, but found the differences negligible.

\change{Other factors that might affect the Ly$\alpha$ transmission, such as the star formation prescription and the quasar halo mass, were investigated in the ATON simulation by \cite{keating_probing_2015} and indicate that the ionized bubble sizes are likely the dominant cause of the damping wing strength between the two simulations employed here.}

Overall, performing two independent analyses using two very different simulations helps us mitigate systematics associated with modelling, thus increasing the robustness of the resultant constraints. At the same time, it serves as a consistency check for reionization simulations, however, a more detailed comparison is beyond the scope of this paper.

\section{Inferring \text{$\vxhi$} and \text{$\logtq$} from stacked quasar spectra}\label{sec:inference}

In this section, we explain the construction of models from the two hydrodynamical simulations and how they are used to infer $\vxhi$ and $\logtq$ for each stacked quasar spectrum.

\subsection{Forward modeling of uncertainties and model construction}\label{sec:noise}

In order to construct realistic models for a grid of $\vxhi$ and $\logtq$ values, we forward-model four different sources of uncertainties present in our observed spectral transmission profiles: the uncertainty from the spectral measurement (coming from the reduction pipeline), the uncertainty from the continuum prediction (originating from QSANNdRA), the uncertainty in quasar redshifts, and cosmic variance (due to the limited number of sightlines in our analysis).

\begin{figure*}
    \centering
    \includegraphics[width=0.45\linewidth]{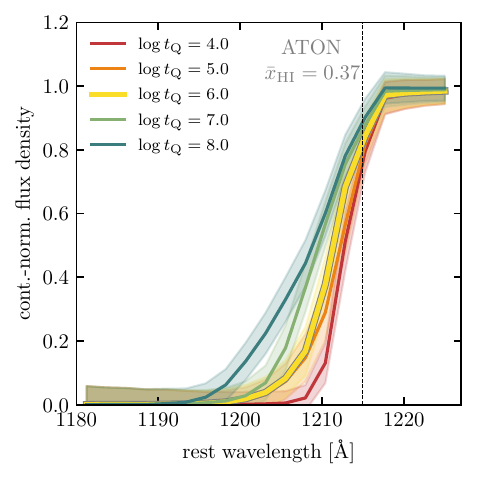}
    \includegraphics[width=0.45\linewidth]{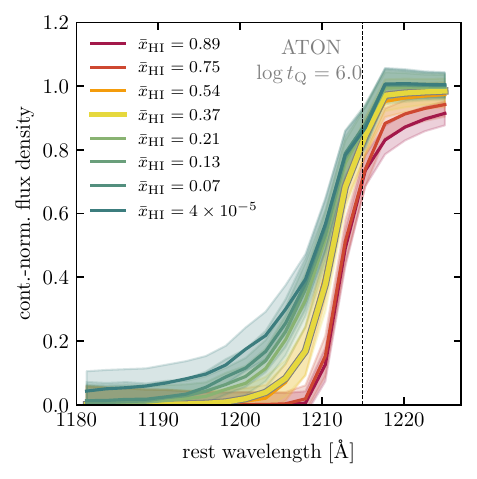}
    \includegraphics[width=0.45\linewidth]{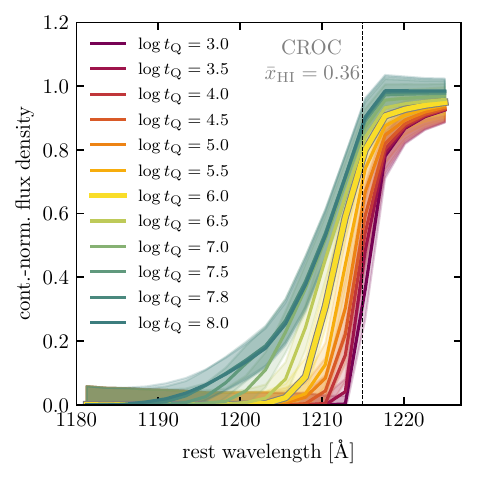}
    \includegraphics[width=0.45\linewidth]{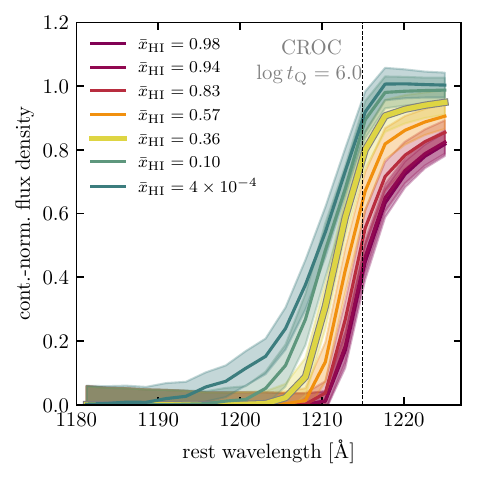}
    \caption{Example models from both ATON (top) and CROC (bottom) for a fixed neutral gas fraction (left) and a fixed quasar lifetime (right). The vertical black dashed line marks the position of \lya. To ease visual comparison, we have highlighted the most comparable model ($\vxhi = 0.37$ (ATON) and $\vxhi = 0.36$ (CROC) with $\logtq = 6.0$) in a thicker line in all four panels.}
    \label{fig:models}
\end{figure*}

In order to account for cosmic variance, we stack the simulation sightlines corresponding to different quasars in the same way as we do with the observed spectra -- we illustrate this forward modeling in \cref{fig:noise}. Specifically, we sample one simulation sightline per quasar in the corresponding redshift bin (leftmost panel of \cref{fig:noise}). For this simulation sightline, we randomly draw a continuum prediction sample from QSANNdRA (second panel from the left in \cref{fig:noise}) as well as a noise vector from a Gaussian distribution centered at 0 and with a standard deviation defined by the measurement \change{uncertainty} at each spectral pixel (third panel of \cref{fig:noise}). This combination of \change{uncertainties} ensures that we model the randomness of the measurement error while still accounting for the fact that the error in continuum prediction is correlated between neighbouring pixels (i.e. it is smoothly varying over wavelength). Afterwards, we sample a new quasar redshift from a Gaussian distribution defined by the quasar redshift $z$ and the corresponding redshift uncertainty $\sigma_z$ as given in \cref{tab:qso_table}, and we shift the sightline wavelength accordingly (fourth panel of \cref{fig:noise}). Having added the first three sources of uncertainty to each quasar sightline, we stack the resultant sightlines to form a sample stacked sightline for each redshift bin (rightmost panel of \cref{fig:noise}\change{, where we also show the the same sample stack without any added uncertainties in gray for comparison)}. At this stage, we also simultaneously rebin the model sightline to the same resolution as our spectral stacks, i.e. $600\ {\rm km/s}$. We repeat this procedure $10000$ times to create $10000$ sample spectral stacks for each redshift bin and for each combination of $\vxhi$ and $\logtq$ provided by the two simulations. The mean of these sample stacks constitutes our model, $\textbf{m} (\vxhi,\logtq,\zbar)$. Note the dependence on $\zbar$ -- each redshift bin has a separate set of models. We show examples of models constructed this way in \cref{fig:models} for both ATON and CROC models (top and bottom panels, respectively). \change{Note that the CROC models on average display stronger damping wings, which is likely due to the smaller ionized bubble sizes allowed by the smaller simulation box, as discussed in Section~\ref{sec:comp}.}

\subsection{Maximum likelihood inference}

\begin{figure}
        \centering
    \includegraphics[width=\linewidth]{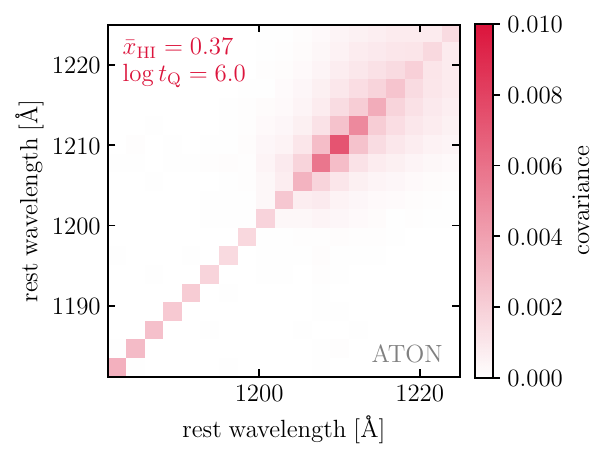}
    \includegraphics[width=\linewidth]{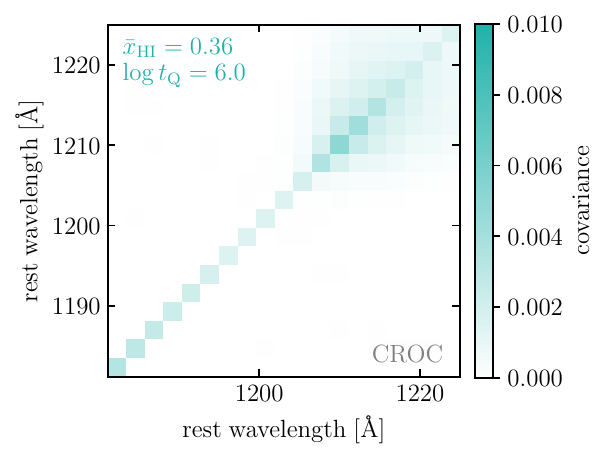}
    \caption{Example covariance matrix for each simulation. Here we show the covariance corresponding to $\vxhi=0.37$ and $\vxhi=0.38$, and $\logtq=6.0$ for ATON (top) and CROC (bottom), respectively. Despite originating from different simulations, the covariance matrices are remarkably similar.}
    \label{fig:cov}
\end{figure}

With models, $\textbf{m} (\vxhi,\logtq,\zbar)$, at hand, we use maximum likelihood estimation to find a combination of $\vxhi$ and $\logtq$ that best fits each of our stacked spectra. We use a multivariate Gaussian log-likelihood function, i.e.
\begin{equation}
    \log \mathcal{L} = -\frac{1}{2} \log \det{\textbf{C}} - \frac{1}{2} (\textbf{y} - \textbf{m})^\mathrm{T} \textbf{C}^{-1} (\textbf{y} - \textbf{m}),
\end{equation}
where $\mathrm{y} = \textbf{y}(\zbar)$ is the observed stacked spectrum at the mean redshift $\zbar$, $\mathrm{m} = \textbf{m}(\vxhi,\logtq,\zbar)$ is the model stacked spectrum, and $\textbf{C} = \textbf{C}(\vxhi,\logtq,\zbar)$ is the covariance matrix computed from the $10000$ stacked spectra samples for each model. The use of covariance matrices allows us to obtain unbiased constraints on $\vxhi$ and $\logtq$ as the spectral flux is invariably correlated across neighboring pixels. Note that we compute the covariance matrices by forward-modeling noise in the models, as each stacked spectrum contains too few sightlines to calculate covariances on the data. We show an example of a covariance matrix at $\vxhi=0.37(0.38)$ and $\logtq = 6.0$ for both ATON and CROC models in \cref{fig:cov}.

\begin{figure*}
    \centering
    \includegraphics[width=0.88\linewidth]{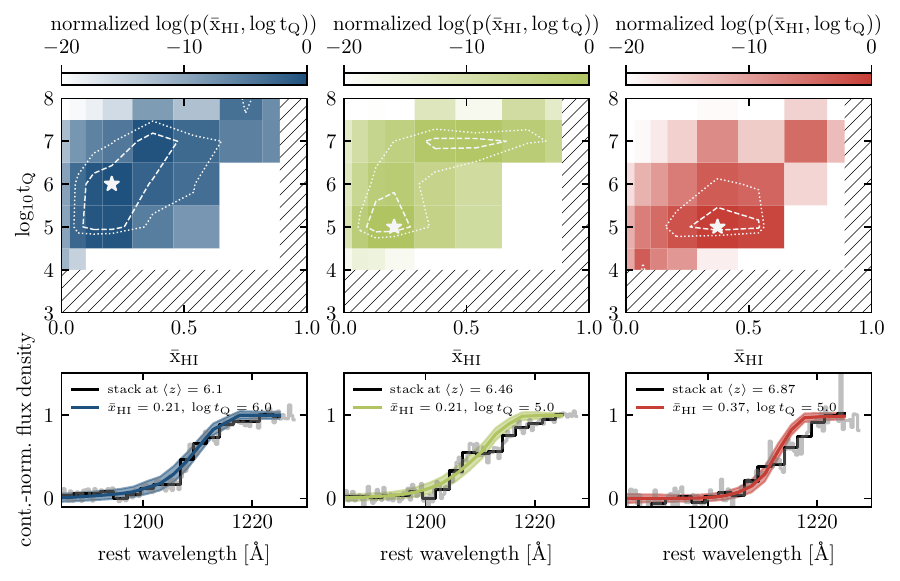}
    \includegraphics[width=0.88\linewidth]{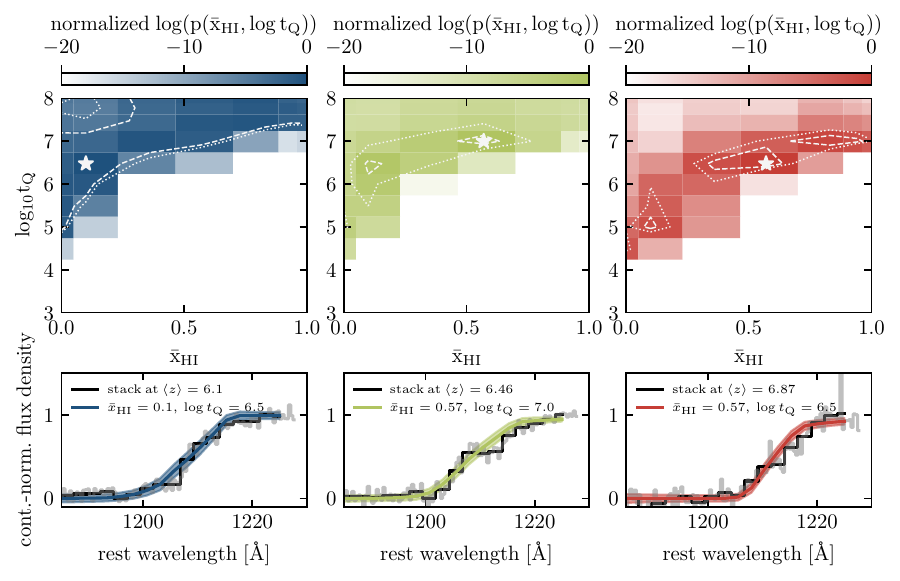}
    \caption{Maximum likelihood estimation with ATON models (top two rows) and CROC models (bottom two rows) for all three redshift bins. In the top panel, we show the 2D joint probability distribution evaluated for each combination of $\logtq$ and $\vxhi$, as well as the $1\sigma$ and $2\sigma$ contours as white dashed and dotted curves, respectively. Note that the parameter space where we do not have ATON models is hatched. In the bottom panel, we plot stacked spectrum for each bin along with the model corresponding to the maximum likelihood point on the 2D grid in the top panel (marked by a star).}
    \label{fig:fits}
\end{figure*}

To infer the neutral fraction and quasar lifetime constraints, we evaluate the log-likelihood function across the parameter grids corresponding to the two simulation models. We calculate the joint probability $p(\vxhi, \logtq)$ by normalizing the likelihood such that $\sum_{\vxhi, \logtq} \mathcal{L}(\vxhi, \logtq) = \sum_{\vxhi, \logtq} p(\vxhi, \logtq) = 1$, and we plot this probability normalized to its peak in \cref{fig:fits,fig:fits} for each of the three aforementioned redshift-stacked spectra using ATON and CROC models, respectively. The peak of $p(\vxhi, \logtq)$ is marked by a star and corresponds to the best fit model shown in the bottom panels of \cref{fig:fits,fig:fits}. We also tested the performance of this method on mock spectral stacks with models and covariances from the respective simulations in \S~\ref{app:mocktest}.

Further, we calculate the posteriors $p(\vxhi)$ and $p(\logtq)$ as
\begin{equation}
    p(\vxhi) \propto \Sigma_{\logtq} \mathcal{L}(\vxhi, \logtq),
\end{equation}
\begin{equation}
    p(\logtq) \propto \Sigma_{\vxhi} \mathcal{L}(\vxhi, \logtq),
\end{equation}
such that $\sum_{\vxhi} p(\vxhi) = 1$ and $\sum_{\logtq} p(\logtq) = 1$, from which we infer the $1\sigma$ uncertainties on our constraints. Due to the coarseness of our parameter space combined with the degeneracy between $\vxhi$ and $\logtq$, the posterior distribution and hence the uncertainties may be asymmetric.

For comparison, we also include 1D inferences of the neutral gas fraction in Appendix \ref{app:fixtq} for both ATON and CROC simulations. There, we assume a fixed mean quasar lifetime of $\logtq = 6.0$ in each redshift bin, a value motivated by the measurement of the average quasar lifetime for a sample of 15 quasars at $z\sim 6$ by \citet{morey_estimating_2021}. Such inference is simpler as it bypasses the problem of the neutral fraction/quasar lifetime degeneracy and thus serves as a consistency check.

Note that not all quasars in each redshift bin are expected to have the same lifetime -- in fact it has been shown that quasar lifetimes at $z\sim3-4$ can be modeled well with a lognormal distribution \citep{khrykin_first_2021}. Our model construction draws only sightlines corresponding to the same quasar lifetime for a particular value of $\logtq$ and hence potentially reduces the underlying variance. We tested our inference pipeline against this concern in \S~\ref{app:lognormaltq}, where we show that our models perform considerably well even on mock spectral stacks with an underlying distribution of lifetimes. In order to avoid additional computational costs, we decided not to relax this assumption.

\section{Results}\label{sec:results}

In this section we present our constraints on the volume-averaged neutral gas fraction, as well as the sample-averaged quasar lifetimes for each redshift bin. The results are summarized in \cref{tab:results}.

\subsection{Constraints on Reionization}\label{sec:reionization}

By marginalizing the joint probability over the grid of quasar lifetimes, we obtain the following volume-averaged neutral fraction constraints in the three redshift bins (\cref{tab:results}): $\vxhi = 0.21_{-0.07}^{+0.17}$ at $\zbar =6.10$, $\vxhi = 0.21_{-0.07}^{+0.33}$ at $\zbar =6.46$, and $\vxhi = 0.37_{-0.17}^{+0.17}$ at $\zbar =6.87$ (ATON, top two rows in \cref{fig:fits}), and $\vxhi = 0.10_{<10^{-4}}^{+0.73}$ at $\zbar =6.10$, $\vxhi =0.57_{-0.47}^{+0.26}$ at $\zbar =6.46$, and $\vxhi =0.57_{-0.21}^{+0.26}$ at $\zbar =6.87$ (CROC, bottom two rows in \cref{fig:fits}). The two different simulations produce consistent reionization constraints. 

\begin{figure*}
    \centering
    \includegraphics[width=\linewidth]{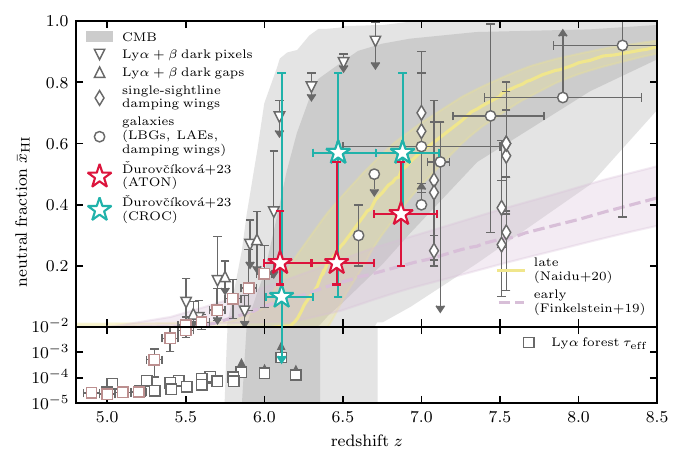}
    \caption{Constraints on the neutral fraction evolution across cosmic time. The constraints from this work are shown in red and blue for ATON and CROC models, respectively. All existing constraints are shown in gray and are overplotted on constraints from the Cosmic Microwave Background \citep[CMB;][]{planck_collaboration_planck_2020}. On the lower redshift end, squares denote constraints from the \lya forest optical depth \change{\citep{fan_constraining_2006,yang_measurements_2020,bosman_comparison_2021,gaikwad_measuring_2023}}, and downward and upward triangles label constraints from \lya and \lyb dark pixels \citep{mcgreer_model-independent_2015,jin_nearly_2023} and dark gaps \citep{zhu_long_2022}, respectively. \change{Note that we highlight the \cite{gaikwad_measuring_2023} constraints in faded red as these are also based on modelling from the ATON simulation and provide a good consistency check for our ATON-based constraints.} At higher redshifts, single-sightline quasar damping wing results are shown as diamonds \citep{wang_luminous_2021,greig_ly_2017,greig_constraints_2019,durovcikova_reionization_2020,davies_quantitative_2018,banados_800-million-solar-mass_2018,yang_poniuaena_2020}, and galaxy constraints from Lyman break galaxies (LBGs), \lya emitters (LAEs) and the recent galaxy damping wing measurements are shown as circles \change{\citep{mason_universe_2018,ouchi_statistics_2010,sobacchi_clustering_2015,mason_inferences_2019,ning_magellan_2022,umeda_jwst_2023}}. We also show two examples of reionization models: one that reionizes late and rapidly \citep[e.g.][]{naidu_rapid_2020}, and one with an earlier and slower reionization \citep[e.g.][]{finkelstein_conditions_2019}. 
    Note that the ATON and CROC constraints are slightly offset in redshift for better visibility.}
    \label{fig:reionization}
\end{figure*}

The maximum likelihood fits of the two higher redshift bins in ATON (green and red in the top of \cref{fig:fits}) seem to underestimate the strength of the observed damping wings. This is simply due to the coarseness of our parameter grid, whereby a neighboring model with a higher neutral gas fraction simply does not provide a better fit than the one shown in \cref{fig:fits}. Also note that our constraints do not directly follow the differences in the damping wings in \cref{fig:zstacks} -- for example, the $\zbar = 6.10$ and the $\zbar = 6.46$ ATON constraints both yield the same maximum likelihood neutral gas fraction despite the differences in the transmitted flux redward of \lya. This is again due to the coarseness of our grid as well as due to the fact that we are jointly fitting for $\vxhi$ as well as the quasar lifetime. This limitation could be overcome by only fitting the red-side damping wings of our spectral stacks, as suggested by \citet{chen_characteristic_2023}, which are agnostic to the stochasticity of the density fields along the individual quasar sightlines -- this is beyond the scope of this paper and will be explored in a separate work.

In some cases, particularly in the case of the lowest two redshift bins (blue and green), the error bars on the CROC constraints are much larger than the error bars on the ATON constraints. In fact, the $1\sigma$ of the CROC posterior reaches the lowest neutral gas fraction in our grid in the $\zbar = 6.10$ case, which is why we indicate these constraints as limits at $\vxhi\sim 10^{-4}$. We hypothesize that the difference in constraining power could be due to the limited number of simulated sightlines (up to $100$ for each snapshot compared with $>500$ per snapshot in ATON), as we do not see such large difference in the highest redshift bin (red). If this were true, the models as well as the covariances would be dominated by the cosmic variance in the simulation and too noisy to provide a precise constraint. The situation is further complicated by the insensitivity to varying neutral gas fractions for the longest simulated lifetimes, which arises from the degeneracy between $\vxhi$ and $\logtq$ and is difficult to mitigate.

It is also worth noting that the $\zbar = 6.46$ likelihood map as well as the posteriors exhibit \change{an apparent} bimodality that is consistent between ATON and CROC -- a good fit is either provided by a longer lifetime quasar embedded in a more neutral IGM or a shorter lifetime quasar in a less neutral IGM. This degeneracy causes the inferred error bars in \cref{fig:reionization} to be asymmetric, and, even more importantly, explains the apparent tension between the two constraints at $\zbar = 6.46$ -- the maximum likelihood models of CROC and ATON, respectively, fall at the two different peaks of this bimodality. The bimodal posterior could be either caused by having two different populations of quasars in our intermediate redshift bin (therefore seeing a combination of small and large proximity zones in the stacked spectrum), \change{however, it is most likely a consequence of the coarseness of our parameter grid -- notably, the $2\sigma$ contours are continuous and do not show the same bimodality as the $1\sigma$ contours. This is especially likely in light of the expected degeneracy between the neutral gas fraction and the quasar lifetime, as is nicely shown in \cite{davies_quantitative_2018}. Therefore, increasing the resolution of our parameter grid for the two simulations, as well as increasing the sample of quasars at this intermediate redshift bin, is expected to reconcile this discrepancy.}

Note that the $\zbar = 6.87$ bin also exhibits a multimodal likelihood map for the CROC models -- in this case the apparent multimodality is \change{again} likely a consequence of the grid coarseness.

In \cref{fig:reionization}, we show a comparison of our neutral gas fraction constraints to existing constraints in the literature. In the lower end of our redshift range, $z\sim 6$, our constraints are consistent with limits from \lya and \lyb dark pixels \cite{mcgreer_model-independent_2015,jin_nearly_2023} and dark gaps \citep{zhu_long_2022}, as well as with limits on the \lya forest opacity \change{\citep{fan_constraining_2006,yang_measurements_2020,bosman_comparison_2021,gaikwad_measuring_2023}. Particularly, note the ATON-based opacity constraints by \cite{gaikwad_measuring_2023} highlighted as faded red squares -- these provide a good consistency check and agree with our ATON-based damping wing modeling in the lowest-redshift bin.} Towards redshifts above $z\gtrsim 7$, we are consistent with constraints from single-sightline quasar damping wings \citep{wang_luminous_2021,greig_ly_2017,greig_constraints_2019,durovcikova_reionization_2020,davies_quantitative_2018,banados_800-million-solar-mass_2018,yang_poniuaena_2020}, as well as with galaxy limits  \citep{sobacchi_clustering_2015,mason_universe_2018,umeda_jwst_2023}. Furthermore, our constraints are also consistent with the dark pixel limits from \cite{jin_nearly_2023} at $6 \lesssim z \lesssim 7$. Overall, the neutral gas fraction constraints in this work bridge the gap between redshifts $z \sim 6$ and $z \sim 7$ and complete the chronicle of the Epoch of Reionization.

Having measurements where previously only limits were present, we now discuss the impact our results have on reionization models. Although there is a growing consensus that the EoR was driven by star-forming galaxies as opposed to quasars \citep{robertson_cosmic_2015,robertson_galaxy_2022}, the population of galaxies that drove reionization remains disagreed upon. Models driven primarily by the most massive galaxies tend to yield late and rapid reionization \citep{naidu_rapid_2020}, whereas a slower, earlier reionization has been demonstrated if the much more abundant, fainter galaxies dominate the ionizing photon budget \citep{finkelstein_conditions_2019,rosdahl_lyc_2022}. However, the implication of late vs. early reionization on the population of ionizing sources rests on many assumptions. For instance, the dependence of the \lya\ escape fraction on the galaxy mass varies between models and can lead to late reionization that is not necessarily driven by the most massive galaxies \citep{keating_long_2020,kulkarni_large_2019}.
Our measurements, as shown in \cref{fig:reionization}, are consistent with both of these alternatives, but slightly favour a late reionization history. At present, we are not able to rule out the early model due to the large uncertainties on our constraints, but the extension of this analysis to a larger sample of quasars \change{in the same luminosity-redshift space while reducing the coarseness of our parameter space grid} should be able to yield a more conclusive answer.

\subsection{Quasar lifetime constraints}

We further marginalize the joint probability over the grid of neutral gas fractions and obtain the following average quasar lifetime constraints in the three redshift bins (\cref{tab:results}): $\logtq = 6.0_{-1.0}^{+1.0}$ at $\zbar =6.10$, $\logtq = 5.0_{-1.0}^{+2.0}$ at $\zbar =6.46$, and $\logtq = 5.0_{-1.0}^{+1.0}$ at $\zbar =6.87$ (ATON, top two rows in \cref{fig:fits}), and $\logtq =6.5_{-0.5}^{+1.3}$ at $\zbar =6.10$, $\logtq =7.0_{-0.5}^{+0.5}$ at $\zbar =6.46$, and $\logtq =6.5_{-0.5}^{+0.5}$ at $\zbar =6.87$ (CROC, bottom two rows in \cref{fig:fits}).

\begin{figure}
    \centering
    \includegraphics[width=\linewidth]{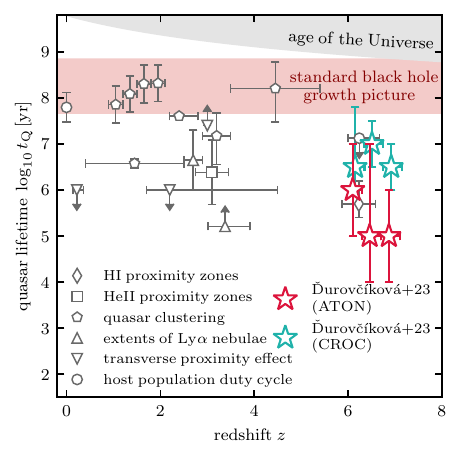}
    \caption{Constraints on quasar lifetimes across redshifts. The measurements from this work are shown in red and blue for ATON and CROC models, respectively. The figure is adapted from \citep{eilers_detecting_2021}, but here we only include datapoints that represent population averages for clarity (all in gray). Constraints from different methods are encoded by symbol: hydrogen and helium proximity zones are shown as diamonds \citep{morey_estimating_2021} and squares \citep{khrykin_first_2021}, respectively, pentagons denote lifetimes measured from quasar clustering \citep{laurent_clustering_2017,shankar_constraints_2010,shen_clustering_2007,white_clustering_2012}, constraints from the extents of \lya nebulae are shown as upward triangles \citep{trainor_constraints_2013,borisova_constraining_2016}, downward triangles encode measurements of the transverse proximity effect \citep{kirkman_transverse_2008,schmidt_statistical_2017,oppenheimer_flickering_2018}, and host population duty cycle studies are shown as circles \citep{yu_observational_2002,chen_constraints_2018}. The horizontal red-shaded band indicates the range of lifetimes one would expect to measure if the black holes underwent a continuous, Eddington-limited accretion. Note that the ATON and CROC constraints are slightly offset in redshift for better visibility.}
    \label{fig:lifetimes}
\end{figure}

Again, the lifetime constraints resulting from the two different simulations are consistent within error bars, with the CROC constraints indicating systematically longer quasar lifetimes. We again observe the effect of the aforementioned bimodality in the $\zbar=6.46$ bin, where the ATON and CROC results deviate more than in the other two bins. As explained in \S~\ref{sec:reionization}, this either suggests two populations of quasars with different lifetimes or that the cosmic variance is too large in this redshift range -- in either case, a larger quasar sample will provide a clarification.

We also note that most of the error bars on lifetime constraints are limited by the coarseness of our model grids. This is particularly true for the ATON models, where we only sample order-of-magnitude differences -- for example, the confidence intervals on the $\zbar = 6.10$ and $\zbar = 6.89$ constraints are quite conservative for this reason.

These quasar lifetime measurements are consistent with other constraints from the literature, as shown in \cref{fig:lifetimes} (note that only population averages instead of individual lifetime measurements are shown for the sake of clarity). Of particular importance is the diamond data point at $z\sim 6$ which is most comparable to our method and represents the only other HI proximity zone population average in the literature to date \citep{morey_estimating_2021}. Additionally, our quasar sample overlaps with the quasar sample used by \cite{morey_estimating_2021}, and so it is crucial to see our constraints be consistent with this measurement. Furthermore, single-sightline damping wing analyses of quasars at $z \gtrsim 7$ by \cite{davies_quantitative_2018,wang_significantly_2020,yang_poniuaena_2020} all find quasar lifetimes $\logtq \lesssim 6.5$, consistent with both of our highest-redshift bin measurements.

Technically speaking, the exact interpretation of the lifetimes inferred from HI proximity zones depends on the ionization state of the IGM. This is because quasars are unlikely to accrete at a constant rate in a single episode -- in fact, quasars are believed to exhibit a flickering light curve whereby the black hole experiences episodes of high and low/no accretion \citep[e.g.][]{zhou_modeling_2023,hopkins_quasars_2009}. 
For a highly ionized IGM, the proximity zone sizes are sensitive to the episodic lifetime, i.e. the most recent episode of quasar activity, as the imprints of any previous accretion phases would have been washed out by the recombination of hydrogen within the proximity zone during the quasar's ``off''-phase. 
On the other hand, a highly neutral IGM enables the measurement of the integrated lifetime, as the abundance of neutral hydrogen impedes a complete recombination within the proximity zone during the quasar's ``off'' phases \citep{davies_evidence_2019}.
Hence, on the lower redshift end of our quasar sample, we expect to be sensitive to the episodic lifetime, whereas the highest redshift quasar spectra are likely beginning to be sensitive to the integrated lifetime. However, stacking multiple quasar sightlines of varying neutral gas fraction renders distinguishing these two scenarios tricky and so we adopt a light-bulb model (where the episodic and the integrated lifetimes are the same) in order to interpret our results.

Under the assumption of a light-bulb light curve, our measurements provide further support for the existence of a young quasar population at high redshifts \citep{eilers_first_2018,eilers_detecting_2021,morey_estimating_2021} and pose a challenge to the standard models of SMBH growth. The masses of most quasars in our sample have been measured and show a mean mass of $1.5\times 10^9$ \msol\ \citep[Bigwood et al. in prep, also][]{farina_x-shooteralma_2022,mazzucchelli_xqr-30_2023}). If a continuous, Eddington-limited accretion with a radiative efficiency of $\epsilon \sim 0.1$ \citep{soltan_masses_1982, yu_observational_2002} is assumed, it takes almost a billion years to grow a black hole mass of $10^9$ \msol\ from a $100$ \msol\ seed \citep{inayoshi_assembly_2020}, i.e.\ several orders of magnitude longer than our measurements suggest. 

Continuous Eddington-limited accretion is unlikely to occur over epochs of billions of years. As has been recently shown in a self-consistent manner \citep{zhou_modeling_2023}, quasars are expected to undergo flickering light curves whereby the black hole accretion rate and thus the number of ionizing photons that enter the IGM varies. As such, growing the supermassive black holes at the centers of these quasars might not be so problematic towards the end of Reionization, as small proximity zones simply indicate only the last ``on'' phase. For higher neutral gas fractions, however, the story of SMBH growth is still quite unclear. Hence, other growth pathways, such as quasar obscuration or radiatively inefficient accretion need to be invoked, details of which are discussed in previous works \citep{eilers_implications_2017,eilers_first_2018,davies_evidence_2019,satyavolu_need_2023}.

\begin{table}
    \centering
    \begin{tabular}{c c c c c}
        \hline
        Quasar & $\vxhi$ & $\vxhi$ & $\logtq$ & $\logtq$ \\
        stack & ATON & CROC & ATON & CROC \\ \hline
        $\zbar = 6.10$ & $0.21_{-0.07}^{+0.17}$ & $0.10_{<10^{-4}}^{+0.73}$ & $6.0_{-1.0}^{+1.0}$ & $6.5_{-0.5}^{+1.3}$ \\ 
        $\zbar = 6.46$ & $0.21_{-0.07}^{+0.33}$ & $0.57_{-0.47}^{+0.26}$ & $5.0_{-1.0}^{+2.0}$ & $7.0_{-0.5}^{+0.5}$ \\ 
        $\zbar = 6.87$ & $0.37_{-0.17}^{+0.17}$ & $0.57_{-0.21}^{+0.26}$ & $5.0_{-1.0}^{+1.0}$ & $6.5_{-0.5}^{+0.5}$ \\ \hline
    \end{tabular}
    \caption{Constraints on the neutral gas fraction, $\vxhi$, and the mean quasar lifetime, $\logtq$, for each of the three redshift bins. The quoted uncertainties are the $16$th and $84$th percentiles of the marginalized posterior. Due to the discreteness of our parameter grid, we quote the distance to the nearest grid value in cases where either percentile coincides with the constraint value -- in this case the uncertainties are not properly resolved by our grid and the $1\sigma$ constraints are therefore conservative.}\label{tab:results}
\end{table}

\section{Summary}\label{sec:summary}

In summary, we have used a sample of $18$ quasars at redshifts $6.0 \leq z \leq 7.1$ and computed continuum-normalized spectral stacks to constrain the neutral gas fraction as well as the mean quasar lifetimes in three redshift bins. 

\begin{itemize}
    \item For the first time, we showcase the emergence of damping wings between redshifts $6 \lesssim z \lesssim 7$ in stacked quasar spectra (\cref{fig:zstacks}).
    \item We constrain the evolution of the volume-averaged neutral gas fraction as follows: 
    $\vxhi = 0.21_{-0.07}^{+0.17}$ at $\zbar =6.10$, $\vxhi = 0.21_{-0.07}^{+0.33}$ at $\zbar =6.46$, and $\vxhi = 0.37_{-0.17}^{+0.17}$ at $\zbar =6.87$ (ATON, top two rows in \cref{fig:fits}), and $\vxhi = 0.10_{<10^{-4}}^{+0.73}$ at $\zbar =6.10$, $\vxhi =0.57_{-0.47}^{+0.26}$ at $\zbar =6.46$, and $\vxhi =0.57_{-0.21}^{+0.26}$ at $\zbar =6.87$ (CROC, bottom two rows in \cref{fig:fits}). These constraints slightly favor late reionization models \citep[e.g.][]{naidu_rapid_2020}, although they are also consistent within the uncertainties with models that reionize early \citep{finkelstein_conditions_2019}, see \cref{fig:reionization}.
    \item We further constrain the mean quasar lifetime in each redshift bin as follows: $\logtq = 6.0_{-1.0}^{+1.0}$ at $\zbar =6.10$, $\logtq = 5.0_{-1.0}^{+2.0}$ at $\zbar =6.46$, and $\logtq = 5.0_{-1.0}^{+1.0}$ at $\zbar =6.87$ (ATON, top two rows in \cref{fig:fits}), and $\logtq =6.5_{-0.5}^{+1.3}$ at $\zbar =6.10$, $\logtq =7.0_{-0.5}^{+0.5}$ at $\zbar =6.46$, and $\logtq =6.5_{-0.5}^{+0.5}$ at $\zbar =6.87$ (CROC, bottom two rows in \cref{fig:fits}). These measurements support the existence of young quasars at $z \gtrsim 6$ (\cref{fig:lifetimes}) and further strain the standard theory of SMBH growth. Obscured or radiatively inefficient growth need to be invoked in order to explain such short quasar lifetimes \citep{eilers_implications_2017,eilers_first_2018,davies_evidence_2019,satyavolu_need_2023}.
    \item Our measurements of the neutral gas fraction and quasar lifetimes are consistent between the two simulations used in this work, ATON and CROC, despite their differences. This serves as an important consistency check for simulation based modeling and increases the robustness of our results.
\end{itemize}

\section*{Data release}

The quasar spectra used in this work will be made public upon the acceptance of this paper. We publish the reduced spectral files along with the smoothed spectral fits and the mean continuum predictions computed in this work (see \S~\ref{sec:cont} for details). See \cref{tab:zenodo} for a description of the published files.

\begin{table*}
    \centering
    \begin{tabular}{c l c l}
        Extension & Contents & Keys & \\
        \hline
        0 & Primary HDU & \\
        1 & Reduced data & \texttt{wave} & observed wavelength $[{\rm \AA}]$\\
         & & \texttt{flux} & flux density $F_\lambda$ $[10^{-17} {\rm ergs/s/cm}^2/{\rm \AA} ]$ \\
         & & \texttt{ivar} & inverse variance of $F_\lambda$ \\
         & & \texttt{mask} & PypeIt mask \\
        2 & Quasar information & \texttt{redshift} & quasar redshift \\
         & & \texttt{redshift\textunderscore err} & redshift error \\
         & & \texttt{M1450} & $M_{1450}$ \\
         & & \texttt{continuum\textunderscore norm} & flux density normalization at $1290\ {\rm \AA}$ $[10^{-17} {\rm ergs/s/cm}^2/{\rm \AA} ]$ \\
        3 & Smoothed red-side spectrum & \texttt{wave\textunderscore red} & red-side rest wavelength $[{\rm \AA}]$ \\
         & & \texttt{flux\textunderscore red} & red-side flux density normalized at $1290\ {\rm \AA}$ \\
        4 & Mean continuum prediction & \texttt{wave\textunderscore blue} & blue-side rest wavelength $[{\rm \AA}]$ \\
         & & \texttt{flux\textunderscore blue} & blue-side flux density normalized at $1290\ {\rm \AA}$ (from QSANNdRA) \\
        5 & 100 prediction samples & \texttt{wave\textunderscore blue\textunderscore 100} & array of blue-side flux densities normalized at $1290\ {\rm \AA}$ \\
        & & & (from QSANNdRA) \\
        \hline
    \end{tabular}
    \caption{Content description of published FITS files for each quasar.}
    \label{tab:zenodo}
\end{table*}

%% Also note that the akcnowlodgment environment does not support long amounts of text. If you have a lot of people and institutions to acknowledge, do not use this command. Instead, create a new \section{Acknowledgments}.

\section*{Acknowledgments}

\change{The authors would like to thank the referee for their helpful comments that helped improve the quality of this manuscript. In addition, the} authors would like to thank Joseph Hennawi, Marianne Vestergaard, Frederick Davies, Sarah Bosman and Molly Wolfson for helpful discussions.

GK is partly supported by the Department of Atomic Energy (Government of India) research project with Project Identification Number RTI 4002, and by the Max Planck Society through a Max Planck Partner Group.

\change{HC thanks the support by the Natural Sciences and Engineering Research Council of Canada (NSERC), funding reference \#DIS-2022-568580.}

We would also like to thank Carlos Contreras, Matías Díaz, Carla Fuentes, Mauricio Martínez, Alberto Pastén, Roger Leiton, Hugo Rivera, and Gabriel Prieto for their help and support during the Magellan/FIRE observations. 

This paper includes data gathered with the 6.5 meter Magellan Telescopes located at Las Campanas Observatory, Chile. 

Based on observations collected at the European Southern Observatory under ESO programmes 084.A-0360, 086.A-0162, 087.A-0607, 098.B-0537, 286.A-5025, 089.A-0814, and 093.A-0707.

Some of the data presented herein were obtained at Keck Observatory, which is a private 501(c)3 non-profit organization operated as a scientific partnership among the California Institute of Technology, the University of California, and the National Aeronautics and Space Administration. The Observatory was made possible by the generous financial support of the W. M. Keck Foundation. The authors wish to recognize and acknowledge the very significant cultural role and reverence that the summit of Maunakea has always had within the Native Hawaiian community. We are most fortunate to have the opportunity to conduct observations from this mountain.

Funding for the Sloan Digital Sky 
Survey IV has been provided by the 
Alfred P. Sloan Foundation, the U.S. 
Department of Energy Office of 
Science, and the Participating 
Institutions. 

SDSS-IV acknowledges support and 
resources from the Center for High 
Performance Computing  at the 
University of Utah. The SDSS 
website is www.sdss4.org.

SDSS-IV is managed by the 
Astrophysical Research Consortium 
for the Participating Institutions 
of the SDSS Collaboration including 
the Brazilian Participation Group, 
the Carnegie Institution for Science, 
Carnegie Mellon University, Center for 
Astrophysics | Harvard \& 
Smithsonian, the Chilean Participation 
Group, the French Participation Group, 
Instituto de Astrof\'isica de 
Canarias, The Johns Hopkins 
University, Kavli Institute for the 
Physics and Mathematics of the 
Universe (IPMU) / University of 
Tokyo, the Korean Participation Group, 
Lawrence Berkeley National Laboratory, 
Leibniz Institut f\"ur Astrophysik 
Potsdam (AIP),  Max-Planck-Institut 
f\"ur Astronomie (MPIA Heidelberg), 
Max-Planck-Institut f\"ur 
Astrophysik (MPA Garching), 
Max-Planck-Institut f\"ur 
Extraterrestrische Physik (MPE), 
National Astronomical Observatories of 
China, New Mexico State University, 
New York University, University of 
Notre Dame, Observat\'ario 
Nacional / MCTI, The Ohio State 
University, Pennsylvania State 
University, Shanghai 
Astronomical Observatory, United 
Kingdom Participation Group, 
Universidad Nacional Aut\'onoma 
de M\'exico, University of Arizona, 
University of Colorado Boulder, 
University of Oxford, University of 
Portsmouth, University of Utah, 
University of Virginia, University 
of Washington, University of 
Wisconsin, Vanderbilt University, 
and Yale University.

For the purpose of open access, the author has applied a Creative Commons Attribution (CC BY) licence to any Author Accepted Manuscript version arising from this submission.


\vspace{5mm}




\appendix

\section{Continuum fits of individual objects}\label{app:continua_zoom}

\begin{figure*}
    \centering
    \includegraphics[width=\linewidth]{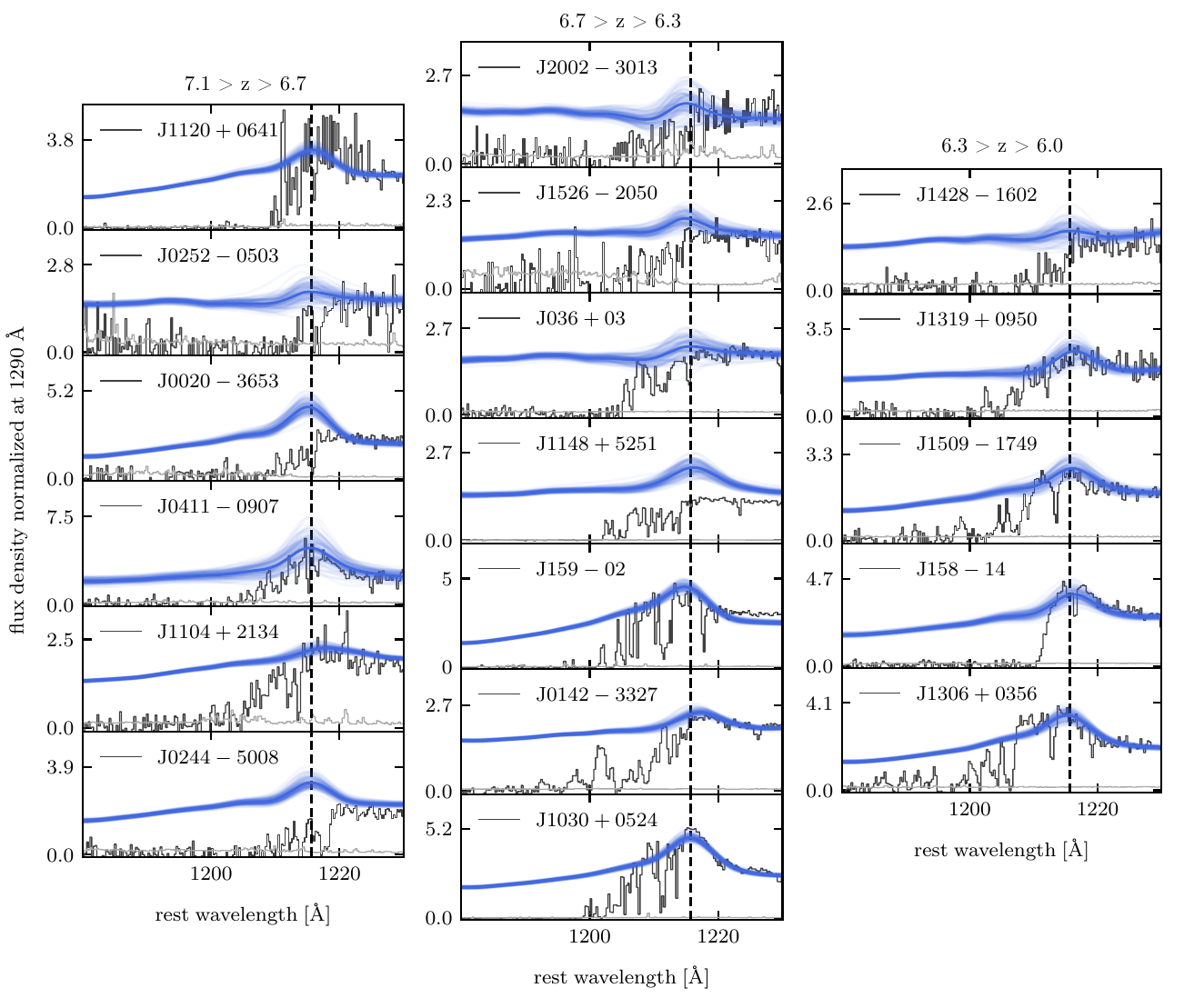}
    \caption{Predicted continua from QSANNdRA (blue) for all quasars in our sample, zoomed in on the wavelength range around the \lya\ emission that is relevant for the inference of $\vxhi$ and $\logtq$. The three columns represent the different redshift bins, from the highest redshift bin on the left to the lowest redshift bin on the right.}
    \label{fig:continua-detail}
\end{figure*}

In \cref{fig:continua-detail}, we show in detail the continuum predictions for all quasars in our sample in the wavelength range that is relevant for the parameter inference described in \S~\ref{sec:inference}.

Of particular interest might be the predicted continua of ULAS J1120+0641, J0252–0503, J1148+5251, and J1030+0524, as the damping wings of these objects have been studied using a variety of methods \citep{mortlock_luminous_2011,davies_predicting_2018,wang_significantly_2020,greig_igm_2022,greig_constraints_2019,schroeder_evidence_2013,mesinger_constraints_2007}, including QSANNdRA \citep{durovcikova_reionization_2020}.

As can be seen in \cref{fig:continua-detail}, QSANNdRA produces a reasonable estimate of the \lya continuum in almost all 18 objects. A few notable exceptions are J159-02 and J1030+0524, where the predicted continuum seems to underestimate the transmitted flux around the \ion{N}{5} line. The interesting thing to note is that in both of these cases, the SDSS nearest neighbor composite shown in \cref{fig:QSANNdRA_predictions} also underpredicts the flux in this region -- this means that either the \ion{C}{3} line that is masked by the region of telluric absorption is unusual in these two quasars, or this discrepancy shows a limitation of the training set of the neutral network. The former explanation would mean that the \ion{C}{3} line profile is important in an accurate reconstruction of the \lya profile. On the other hand, J1148+5251 is predicted to have an exceptionally strong damping wing, in agreement with the SDSS nearest neighbor composite. Upon closer inspection, one can notice that the \ion{C}{4} line is partially masked due to atmospheric absorption, which renders this prediction less reliable due to our inability to make out the full profile of this emission line.

\section{Inference pipeline testing}\label{app:mocktest}

\begin{figure*}
    \centering
    \includegraphics[width=0.7\linewidth]{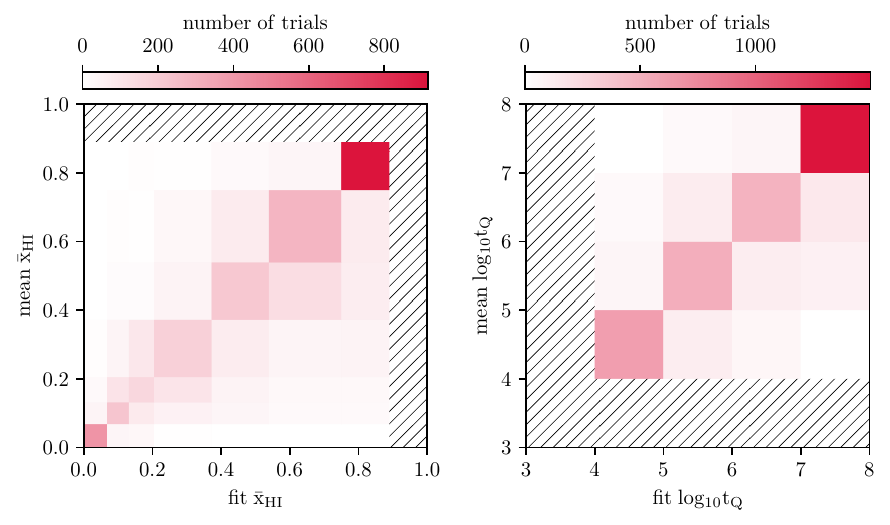}
    \includegraphics[width=0.7\linewidth]{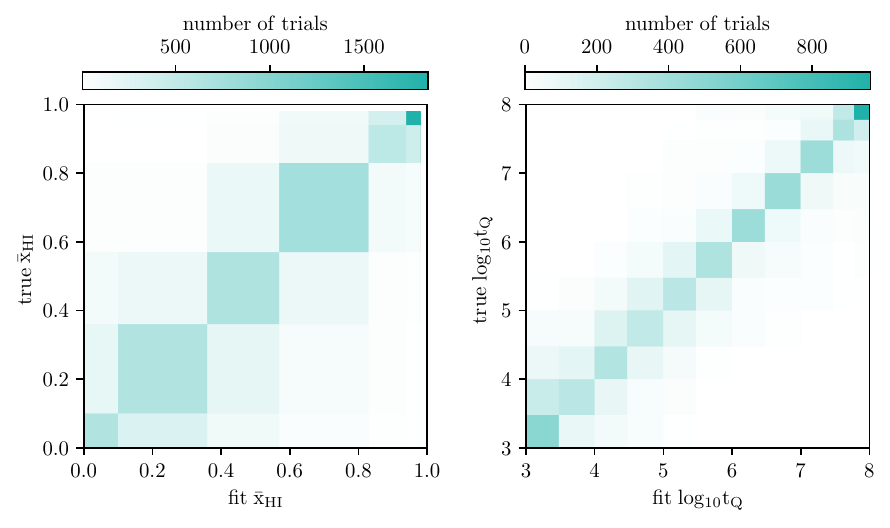}
    \caption{Confusion matrices displaying the performance of our inference pipeline on mock stacked spectra for both ATON (top panels) and CROC (bottom panels) simulations. In all four panels, we plot the true value of $\vxhi$ (left) and $\logtq$ (right) on the y-axis against the respective best fit values on the x-axis. All four confusion matrices are mostly diagonal, with the off-diagonal terms being a consequence of the forward-modelled uncertainties as well as the aforementioned degeneracy.}
    \label{fig:inference_test}
\end{figure*}

In order to ensure that the inference pipeline described in \S~\ref{sec:inference} works as expected, we tested that it works on mock stacked spectra from simulations. To this end, we perform the same procedure as outlined in \S~\ref{sec:noise} for each simulation to create mock spectral samples for each redshift bin -- more specifically, we draw the corresponding number of sightlines for each quasar, add the measurement, continuum prediction and redshift uncertainties, and stack these mock sightlines to produce a sample stacked spectrum at a known neutral fraction, $\vxhi$, and quasar lifetime, $\logtq$. Subsequently, we run our inference pipeline on this mock stacked spectrum to test whether we are able to fit the true $\vxhi$ and $\logtq$. We repeated this procedure $100$ times for each combination of the neutral fraction and quasar lifetime, and we plot the resultant performance as confusion matrices in \cref{fig:inference_test}. 

In the case of both simulations, ATON (top) and CROC (bottom), we see that the resultant confusion matrices are mostly diagonal, with the best accuracy achieved for the highest neutral gas fractions and the longest quasar lifetimes. It is understandable that we see some off-diagonal terms -- these are consequences of the different sources of noise that we are forward modeling as well as the degeneracy between the neutral fraction and quasar lifetimes. Overall, \cref{fig:inference_test} indicates that our inference pipeline works well.

\section{Fitting $\vxhi$ at a fixed lifetime}\label{app:fixtq}

For completeness, here we show the maximum likelihood inferences of the neutral gas fraction assuming a fixed quasar lifetime, $\logtq = 6.0$. The choice of this quasar lifetime is motivated by the first measurement of the mean quasar lifetime in a sample of $z\sim 6$ quasars \citep{morey_estimating_2021}.

\begin{figure*}
    \centering
    \includegraphics[width=\linewidth]{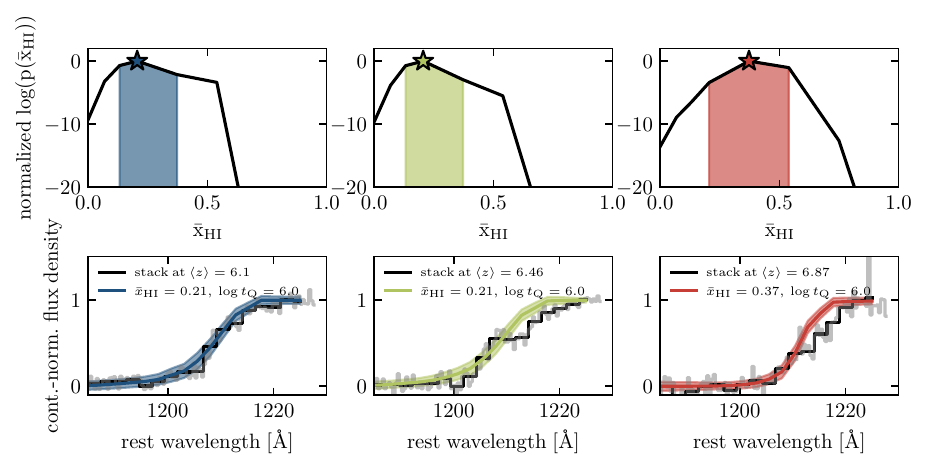}
    \includegraphics[width=\linewidth]{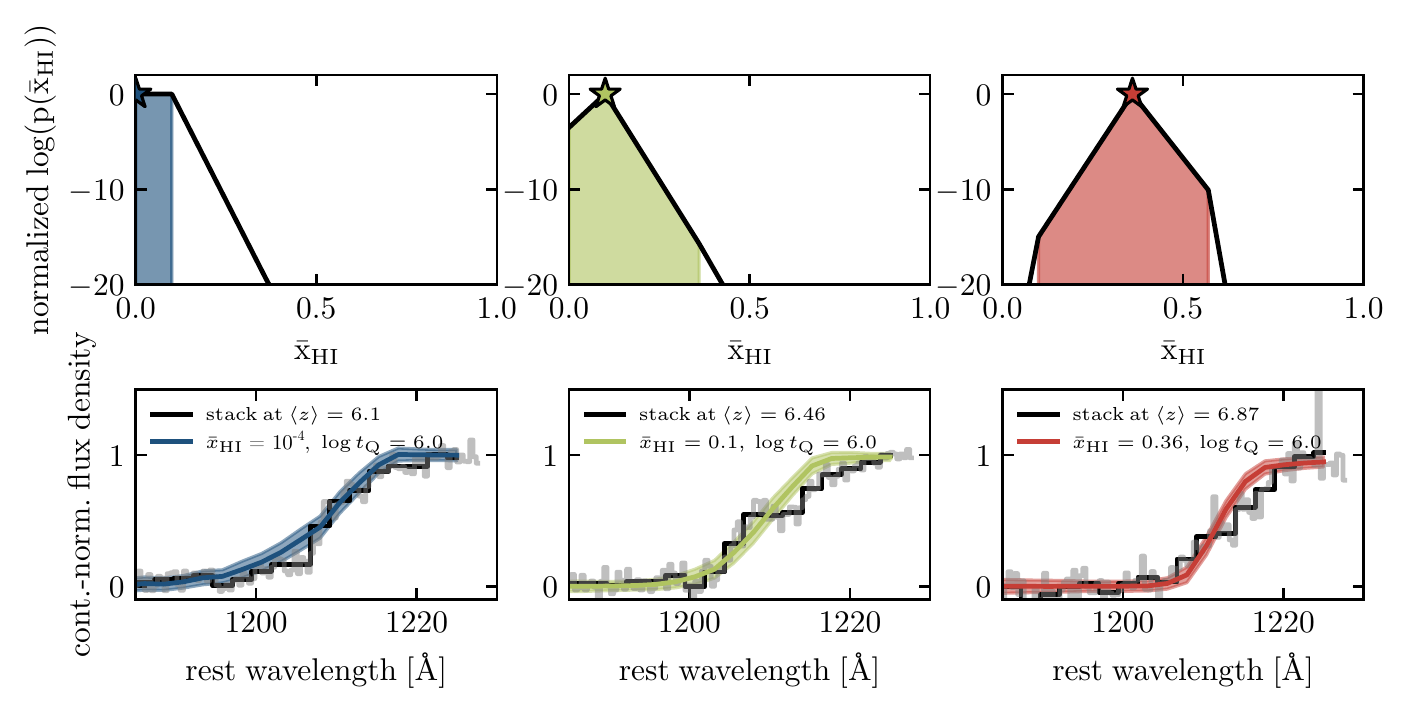}
    \caption{Maximum likelihood inference of the neutral gas fraction, $\vxhi$, for a fixed quasar lifetime, $\logtq = 6.0$. The choice of this lifetime value was inspired by the mean quasar lifetime measured for a sample of $z\sim 6$ quasars \citep{morey_estimating_2021}. The top two rows correspond to the ATON simulation, whereas the bottom two rows correspond to the CROC simulation. The top panels in each case show the log-space normalized posterior probability $p(\vxhi)$, and the stars mark the position of the best fit plotted in the bottom panels.}
    \label{fig:fixtq}
\end{figure*}

The resultant inferences are shown in \cref{fig:fixtq} for both the ATON simulation (top two rows) and the CROC simulation (bottom two rows). By assuming a prior on the mean lifetime in each redshift bin, the inference on the neutral gas fraction becomes simpler -- in particular, we do not observe the posterior bimodality in the $\zbar = 6.46$ that arises in the joint inference in \cref{fig:fits}.

The fits shown in \cref{fig:fixtq} showcase a monotonical neutral gas fraction evolution with redshift, as one would expect based on the visual inspection of the damping wing profiles \cref{fig:zstacks}. However, only estimating the neutral gas fraction does not capture the full picture due to the degeneracy between $\vxhi$ and $\logtq$, which is why we only include this analysis in the appendix and focus on the joint analysis in the main text.

\section{Mean lifetime vs distribution of lifetimes}\label{app:lognormaltq}

In \S~\ref{sec:noise}, where we described the forward modeling of uncertainties and the construction of models, we noted that the simulation sightlines for a model at a fixed $\logtq$ were all drawn from runs corresponding to the same quasar lifetime. This is equivalent to assuming that all quasars in a particular redshift bin have the same lifetime, equal to the mean lifetime that we infer alongside the neutral gas fraction. In this section, we test our models against relaxing this assumption. Specifically, we test whether a mock simulation stack that constains sightlines at a range of lifetimes for a particular $\vxhi$ can be accurately fit with our models for its true neutral gas fraction and its true mean quasar lifetime.

\begin{figure*}
    \centering
    \includegraphics[width=0.7\linewidth]{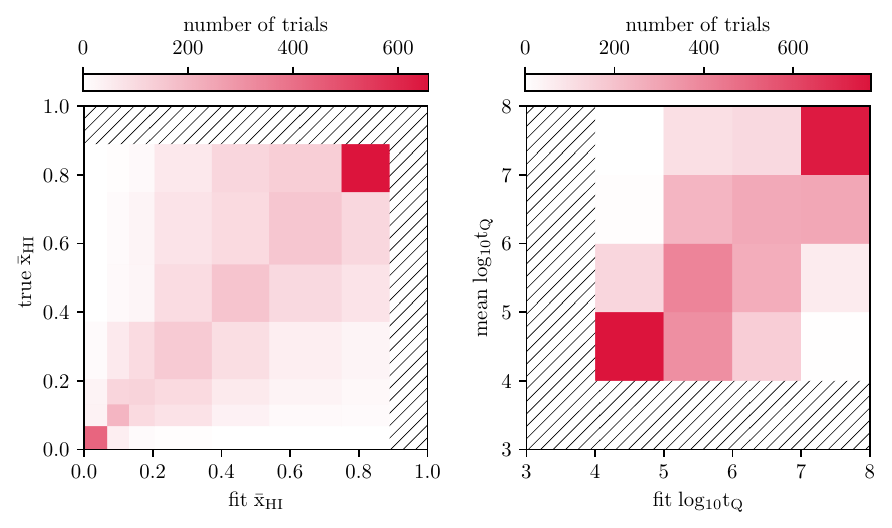}
    \includegraphics[width=0.7\linewidth]{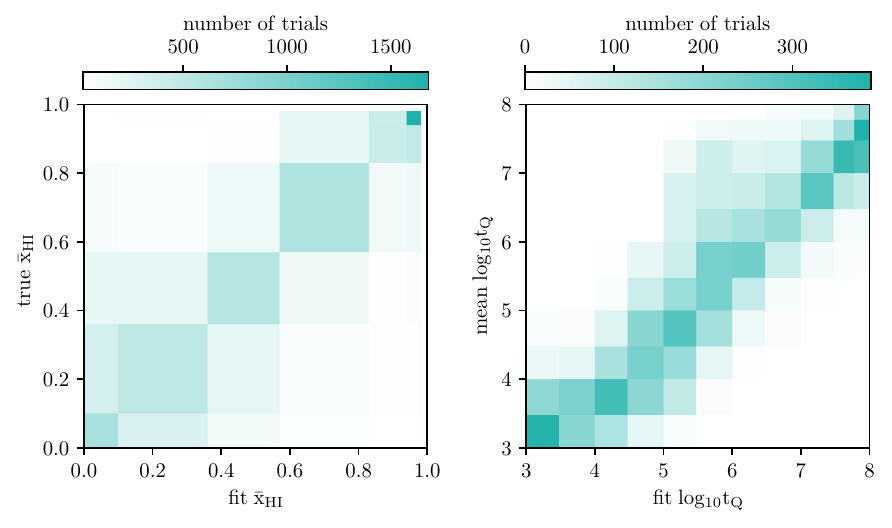}
    \caption{Confusion matrices displaying the performance of our inference pipeline on mock stacked spectra at a range of quasar lifetimes for both ATON (top panels) and CROC (bottom panels) simulations. In all four panels, we plot the true value of $\vxhi$ (left) and the true mean quasar lifetime of the mock stack, $\logtq$, (right) on the y-axis against the respective best fit values on the x-axis. This test demonstrates that our inference of the neutral gas fraction is robust against quasars in our stacked spectra having different lifetimes.}
    \label{fig:lognormaltq}
\end{figure*}

To this end, we follow the same procedure as in \S~\ref{app:mocktest} but for a given $\vxhi$, we stack sightlines corresponding to $N_{\rm QSO}$ draws of $\logtq$ from a lognormal distribution centered on the mean quasar lifetime. Inspired by \citet{khrykin_first_2021}, we assume the variance of the lifetime distributions to be $1$ dex. We plot the resultant confusion matrices in \cref{fig:lognormaltq} for both ATON (top) and CROC (bottom). Note that we keep the same forward modeling of uncertainties as in \S~\ref{app:mocktest}.

The right panels in \cref{fig:lognormaltq} now show a more ``fuzzy'' diagonal structure -- it is now harder for our models to infer the correct mean quasar lifetime in the particular mock stack. Notwithstanding, the resultant confusion matrices for the neutral gas fractions (the left panels) are still fairly diagonal, which showcases that our method recovers the true neutral gas fraction considerably well even if the quasars in our stacked spectra (\cref{fig:zstacks}) have different lifetimes.


\bibliography{fireqsos}{}
\bibliographystyle{aasjournal}



\end{document}